\documentclass[a4paper,aps,prd,twocolumn,eqsecnum,amsmath,amssymb,nofootinbib,superscriptaddress]{revtex4-1}
\usepackage{dcolumn}
\usepackage{bm}
\usepackage{graphics}
\usepackage{graphicx}
\usepackage{epsfig}
\usepackage{booktabs,multirow}
\usepackage{hhline}
\usepackage{slashed}
\usepackage{rccol}
\usepackage{mathrsfs}
\usepackage[dvipsnames]{xcolor,colortbl}

\usepackage{xcolor}\colorlet{linkequation}{blue}
\usepackage[colorlinks=true, % false: boxed links; true: colored links
            linkcolor=red,   % color of internal links
            filecolor=red,   % color of file links
            citecolor=green, % color of links to bibliography
            urlcolor=blue,
            hyperfootnotes=true]{hyperref}

\newcommand*{\SavedEqref}{}
\let\SavedEqref\eqref
\renewcommand*{\eqref}[1]{%
  \begingroup
    \hypersetup{
     linkcolor=linkequation,
      linkbordercolor=linkequation,
    }%
    \SavedEqref{#1}%
  \endgroup
}

\begin{document}

% My Shortcuts
\def\beqa{\begin{eqnarray}}
\def\eeqa{\end{eqnarray}}
\newcommand{\be}{\ensuremath{\beta}}
\newcommand{\al}{\ensuremath{\alpha}}
\newcommand{\sa}{\ensuremath{\sin\alpha}}
\newcommand{\ca}{\ensuremath{\cos\alpha}}
\newcommand{\ta}{\ensuremath{\tan\alpha}}
\newcommand{\sbt}{\ensuremath{\sin\beta}}
\newcommand{\cbt}{\ensuremath{\cos\beta}}
\newcommand{\cba}{\ensuremath{c_{\beta-\alpha}}}
\newcommand{\ma}{\ensuremath{m_{A}}}
\newcommand{\mh}{\ensuremath{m_{h^0}}}
\newcommand{\mH}{\ensuremath{m_{H^0}}}
\newcommand{\mev}{\mbox{~MeV}}
\newcommand{\gev}{\mbox{~GeV}}
\newcommand{\tev}{\mbox{~TeV}}

\newcommand{\ben}{\begin{enumerate}}
\newcommand{\een}{\end{enumerate}}
\newcommand{\bc}{\begin{center}}
\newcommand{\ec}{\end{center}}
\newcommand{\mb}{\mbox{\ }}
\newcommand{\vs}{\vspace}
\newcommand{\ra}{\rightarrow}
\newcommand{\la}{\leftarrow}
\newcommand{\ul}{\underline}
\newcommand{\ds}{\displaystyle}
%\numberwithin{equation}{section}
\definecolor{LightCyan}{rgb}{0.0, 1, 0.94}

\title{One-loop electroweak radiative corrections to charged lepton pair production in photon-photon collisions}

\author{M.~Demirci}
\email{Corresponding author. mehmetdemirci@ktu.edu.tr}
\affiliation{Department of Physics, Karadeniz Technical University, TR61080 Trabzon, Turkey}%
\author{M. F. Mustamin}
\email{mfmustamin@ktu.edu.tr}
\affiliation{Department of Physics, Karadeniz Technical University, TR61080 Trabzon, Turkey}%
\date{\today}
\begin{abstract} We provide high-precision predictions for muon-pair and tau-pair productions in a photon-photon collision by considering a complete set of one-loop-level scattering amplitudes, i.e., electroweak (EW) corrections together with soft and hard QED radiation. Accordingly, we present a detailed numerical discussion with particular emphasis on the pure QED corrections as well as genuinely weak corrections. The effects of angular and initial beam polarisation distributions on production rates are also discussed. An improvement is observed by a factor of two with oppositely polarized photons. Our results indicate that the one-loop EW radiative corrections enhance the Born cross section and the total relative correction is typically about ten percent for both production channels. It appears that the full EW corrections to $\gamma \gamma \to \ell^- \ell^+$ are required to match a percent level accuracy.
\keywords{charged-leptons, EW radiative corrections, standard model, photon-photon collider}
\end{abstract}

\pacs{14.80.Cp, 12.15.Lk, 12.60.Fr, 12.38.Bx}

\maketitle

\section{Introduction}
The Standard Model (SM) of particle physics~\cite{Glashow61,Weinberg67,Salam68} has been perfectly proved as a self-consistent gauge theory with a weakly-coupled sector for electroweak (EW) symmetry breaking with the discovery of a 125 GeV of Higgs boson~\cite{ATLAS,CMS} and the ever-increasing confidence of its compatibility with the SM Higgs boson~\cite{ATLASCMS} at the LHC. With this achievement, further challenges of particle physics lie in expanding the current theory to explain phenomena beyond the SM (BSM) like dark matter, hierarchy problem, strong CP problem and generation of a baryon asymmetry; and improve the accurate measurement of observed phenomena. Advancement of current experiments is still ongoing in pursuing these aims which will offer opportunities to precisely study properties of our current knowledge of fundamental particles.

Being a proton-proton collider, LHC generates plentiful particles so that studying a particular process attached with the abundant background. However, probing phenomenology with another machine, such as electron-positron colliders with a cleaner background, offers significant opportunities. The clean circumstances in these facilities would ensure that the interested phenomena would be precisely observed. International Linear Collider (ILC)~\cite{ILC1,ILC2} is one of the proposed machines for this case. It is aimed to construct equipment for $e^-e^+$ collision, together with other type of collision modes like $e^-e^-$, $e^-\gamma$ and $\gamma\gamma$. Another proposal is the Compact Linear Collider (CLIC)~\cite{CLIC1}, planned to have a TeV–scale high-luminosity working with $\sqrt{s}$ up to $3\tev$. Another prospective clean channel is photon-photon ($\gamma\gamma$) collision. This collision mode could provide an integrated luminosity by the order of 1000 fb$^{-1}$ per year. Advancement of the facility is expected to supply the high energy of $1\tev$ with up to 300 fb$^{-1}$ per year of total integrated luminosity~\cite{ILC3}.

The latter channel provides an interesting framework to improve our understanding of the current theory. High energy $\gamma\gamma$ collisions will play an important role as a comprehensive framework for virtually investigating every aspect of the SM and beyond since photon couples directly to all fundamental fields via electromagnetic current, such as leptons, quarks, $W$'s, and supersymmetric particles. Some recent study regarding this mode have been done for example on describing $\mu^+ \mu^-$ resonance \cite{Godunov2021}, quarkonium pairs with QCD correction \cite{Yang20}, neutralino pair production~\cite{Demirci16}, and also Higgs production \cite{Demirci19a,Demirci20, Enterria20}. The relatively clean environment of this collision, in particular, opens an opportunity for further precision tests of the SM which will complement the ones obtained from future $e^-e^+$ scattering facilities. High-energetic processes of photon-photon collision can be generated through Compton backscattering of laser light off high-energy electrons~\cite{Ginzburg83,Ginzburg84}. Moreover, the elastic backscattering of this kind is well-suited for monitoring the luminosity of such collider in analogy to Bhabha forward scattering in $e^-e^+$ colliders. Lepton pair production in photon-photon collisions, as an implication, represents one of the most important processes in this context.

It is known that the main reactions for the charged leptons pair production in the modern and future $e^-e^+$ colliders are $e^- e^+ \rightarrow \ell^{-}\ell^{+}$ and $\gamma\gamma \rightarrow \ell^{-}\ell^{+}$. Since the cross-section of the former is suppressed by s-channel contributions at high energies, the production rate of the latter can give larger measurements. The process $e^- e^+ \rightarrow \ell^{-}\ell^{+}$ has been intensively investigated by considering the one-loop EW radiative corrections in the SM (see e.g. in Refs.~\cite{Bardin:1981sv,Bardin:1980fe,Akhundov:1984mp,Berends:1987bg,Bardin:1989tq,Hollik:1988ii} before the LEP era, and recently in Refs.~\cite{Bondarenko18,Bondarenko20}). On the other hand, there are only a few works for lepton pair production via the $\gamma\gamma$-collision mode. For instance, the complete $\mathcal{O}(\alpha)$ corrections to  $\gamma\gamma \rightarrow \ell^{-}\ell^{+}$ have been calculated in Ref.~\cite{Denner99} for arbitrary light fermions (only $\ell=e$), where fermion-mass effects are neglected.

Aiming to have accurate measurements, high precision predictions from the model are necessary. Meaning that for most processes there is a need to proceed beyond the lowest order calculations. In particular, a complete set of one-loop corrections is crucial for analyzing phenomena at the future colliders. In this work, we investigate production of the charged lepton pairs via $\gamma\gamma$ collision for heavy leptons $\mu$ and $\tau$ in the framework of SM, including full one-loop EW radiative corrections, i.e. EW corrections, as well as soft and hard QED radiation. We give a detailed numerical discussion accordingly with particular emphasis on the pure QED and weak corrections. In addition, we take into account the decomposition of the weak corrections into the purely fermionic loops along with the fermionic part of the counter-terms and the remaining bosonic corrections. We also present numerical results for angular distribution and initial beam polarisation distribution on the Born-level and one-loop cross sections.

The rest of this work is arranged as follows. In Sec.~\ref{sec:cros}, we present the Feynman diagrams, the relevant amplitudes, and some useful analytical expressions. We also discuss the general shapes of the virtual and the real photon radiation contributions. In Sec.~\ref{sec:results}, we present the numerical evaluations of the radiative corrections related to the scattering process, and discuss in detail relative corrections of QED and Weak. We also present a comparison with the results of other tools. Finally, the concluding remarks are given in Sec.~\ref{sec:conc}.

\section{Theoretical Setup for cross section}\label{sec:cros}

In this section, some details on the calculation of the born-level and the full EW $\mathcal{O}(\alpha)$ corrections are respectively given for the production of charged lepton pairs in $\gamma\gamma$ collision mode. We express the relevant scattering process as
\begin{equation} \label{eq:gammagammaHH}
\gamma(p_1,\lambda_1)\gamma(p_2,\lambda_2)\rightarrow \ell^{+}(k_1,\sigma_1) \ell^{-}(k_2,\sigma_2),
\end{equation}
where $\lambda_1$, $\lambda_2$, $\sigma_1$ and $\sigma_2$ are the helicities of initial photons and final leptons, respectively. The helicities take the values $\lambda_{1,2}=\pm1$ and $\sigma_{1,2}=\pm1/2$. All momenta $p_1,p_2,k_1$ and $k_2$ obey the on-shell equations $p_1^2=p_2^2=0$ and $k_1^2=k_2^2=m_{l^\pm}^2$. For further use, also note the Mandelstam variables:
\begin{equation}
\begin{split}
\hat{s}=(p_1+p_2)^2,
\hat{t}=(k_1-p_1)^2,
\hat{u}=(k_2-p_1)^2.
\end{split} \label{eq:manvar}
\end{equation}
In the center-of-mass system of the final states, we have,
\begin{equation}
\begin{split}
p_1&=\frac{\sqrt{\hat{s}}}{2}(1,0,0,-1),\\
p_2&=\frac{\sqrt{\hat{s}}}{2}(1,0,0,+1),\\
k_1&=\frac{\sqrt{\hat{s}}}{2}(1,-\sin\theta,0,-\cos\theta),\\
k_2&=\frac{\sqrt{\hat{s}}}{2}(1,+\sin\theta,0,+\cos\theta),
\end{split} \label{eq:cms_momentum}
\end{equation}
where $\theta$ denotes the scattering angle.
In order to be used in the calculation of polarized cross sections, we present the photon polarization vectors as
\begin{equation}
\begin{split}
\varepsilon^1_{\mu}(p_1,\lambda_1=\pm1)&=-\frac{1}{\sqrt{2}}(0,1,\mp i,0),\\
\varepsilon^2_{\mu}(p_2,\lambda_2=\pm1)&=\frac{1}{\sqrt{2}}(0,1,\pm i,0),
\end{split}
\end{equation}
which ensure $\varepsilon^{i}\cdot p_j=0$ for $i,j=1,2$. The photon beams circular polarization provides two possible running modes in terms of helicities: a parallel and an anti-parallel alignment of the photon helicities. They are respectively equivalent to the overall angular momentums of $J_z = 2$ and $J_z = 0$. Basically, one can obtain either laser photons or $e^-$-beams with a suitable choice of helicity.

The analytical and numerical evaluations have been made using packages\footnote{
We have already carried out several recent works~\cite{Demirci16, Demirci19a,Demirci19b,Demirci14,Demirci20}, which include significant results, by using the same tools.} described as follows.
The Feynman diagrams and amplitudes have been generated with the means of \textsc{FeynArts}~\cite{Feynarts}. Technically, the algebraic evaluation of the Feynman amplitudes has been performed in the same way as described in Ref.~\cite{Demirci20} for $\gamma\gamma \rightarrow H^{-}H^{+}$. Next, squaring the relevant amplitudes, simplifying the fermion chains, and the numerical computation have been performed with help of \textsc{FormCalc} \cite{loop}. The scalar loop integrals have been computed with the help of \textsc{LoopTools}~\cite{loop}. The phase-space integrations are computed via the Monte-Carlo integration algorithm Vegas, implemented in the \textsc{CUBA} library~\cite{CUBA}. The evaluation of the hard photon bremsstrahlung process has been successfully checked against the results obtained with \textsc{CalcHEP}~\cite{CalcHep} and \textsc{Whizard}~\cite{Whizard,Omega}.

\subsection{Lowest-Order Calculation}
\begin{figure}[hbt]
    \begin{center}
\includegraphics[width=0.80\linewidth]{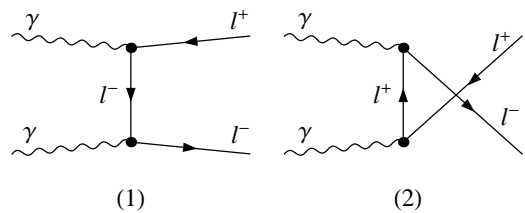}
     \end{center}
          \vspace{-3mm}
\caption{The lowest order Feynman diagrams contributing to $\gamma\gamma \rightarrow \ell^{-}\ell^{+}$.}\label{fig:borndiagram}
\end{figure}

In the lowest order, process $\gamma\gamma \rightarrow \ell^{-}\ell^{+}$ is a pure QED process in which the leading contribution comes from $t$ and $u$-channel charged lepton-exchange diagrams. We present the Born-level Feynman diagrams in Fig.~\ref{fig:borndiagram}.
The matrix elements for these diagrams are
\begin{equation}
\begin{split}
{\cal M}_1=&\frac{ -i e^2}{\big[\hat{t}-m_{\ell}^2\big]}\overline{u}(k_2,m_\ell)\gamma^\nu \varepsilon^{\nu}(p_2)(\slashed{k}_2-\slashed{p}_2+m_\ell) \\
&\times \varepsilon^{\mu}(p_1) \gamma^\mu v(k_1,m_\ell),
\end{split}
\end{equation}

\begin{equation}
\begin{split}
{\cal M}_2=&\frac{ -i e^2}{\big[\hat{u}-m_{\ell}^2\big]}\overline{u}(k_2,m_\ell)\gamma^\mu \varepsilon^{\nu}(p_2) (\slashed{p}_2-\slashed{k}_1+m_\ell)\\
&\times \varepsilon^{\mu}(p_1) \gamma^\nu v(k_1,m_\ell),
\label{eq:m1}
\end{split}
\end{equation}
where $\varepsilon_{\mu}(p_1)$ and $\varepsilon_{\nu}(p_2)$ refer to polarization vector of initial photons, and $\alpha=\frac{e^2}{4\pi}$. The Born-level total amplitude is given by
\begin{equation}\label{eq:totMee}
{\cal M}_{\text{Born}}=\sum_{i=1}^{2} {\cal M}_{i},
\end{equation}
leading to the differential cross section
\begin{equation} \label{eq:difsigma}
\begin{split}
\biggl(\frac{d\hat{\sigma}}{d\Omega }\biggr)_{\text{Born}}^{\gamma\gamma\rightarrow \ell^{-}\ell^{+}} = &\frac
{1}{64 \pi^2 \hat{s}} \sum_{ \lambda_{1,2},\sigma_{1,2}}\frac{1}{4} (1+P_1 \lambda_1)(1+P_2 \lambda_2)\\
& \times |{\cal M}_{\text{Born}}^{\lambda_1,\lambda_2,\sigma_1,\sigma_2}|^2,
\end{split}
\end{equation}
where $P_1$ and $P_2$ are the polarization degree of the incoming photons. Following the square of total amplitude and the summation over final particle helicities, the integrated cross-section is obtained by
\begin{equation} \label{eq:totalsigma}
\hat{\sigma}_{\text{Born}}^{\gamma\gamma\rightarrow \ell^{-}\ell^{+}}=\frac
{1}{16\pi \hat{s}^{2}}\int_{\hat
t^{-}}^{\hat t^{+}} \biggl(\frac{1}{4}\biggr)\sum_{\lambda_{1,2},\sigma_{1,2}} |{\cal M}_{\text{Born}}|^2 d\hat t
\end{equation}
where the factor (1/4) comes from averaging spin of the initial photons and
\begin{equation}
\hat{t}^\pm=(m_{l^\pm}^2-\frac{\hat{s}}{2}) \pm\frac{1}{2}\bigl(\sqrt{\hat{s}^2-4 \hat{s} m_{l^\pm}^2 }\bigr).
\end{equation}

\subsection{One-loop EW Radiative Corrections}

\subsubsection{Virtual corrections}
Higher-order contributions are required to increase precision in the analysis of high energy processes in modern and future colliders. The process~\eqref{eq:gammagammaHH} includes one-loop level contributions in order of $\mathcal{O}(\alpha)$, which based on pure EW corrections. It is known that the total amplitude at one-loop level may be taken as a linear sum of triangle, bubble and box one-loop integrals. Accordingly, the virtual contributions for the process $\gamma \gamma \rightarrow \ell^{-}\ell^{+}$ come from three different types of diagrams, according to the loop correction type: box-type, self-energy, and vertex-type diagrams.

\begin{figure}[hbt]
    \begin{center}
\includegraphics[width=1\linewidth]{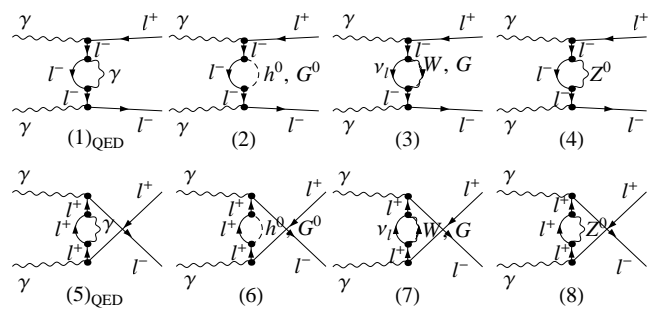}
     \end{center}
     \vspace{-3mm}
\caption{The self-energy diagrams contributing to $\gamma \gamma \rightarrow \ell^{-}\ell^{+}$.}\label{fig:diagself}
\end{figure}

\begin{figure}[hbt]
    \begin{center}
\includegraphics[width=1\linewidth]{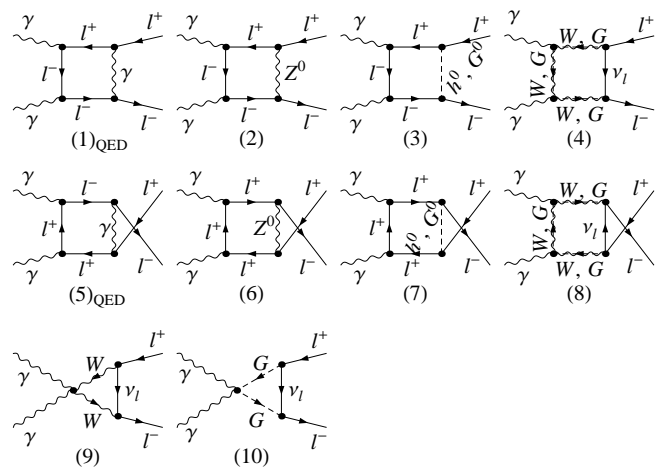}
     \end{center}
          \vspace{-3mm}
\caption{The box-type diagrams contributing to $\gamma \gamma \rightarrow \ell^{-}\ell^{+}$.}\label{fig:diagbox}
\end{figure}

\begin{figure*}[htb]
    \begin{center}
\includegraphics[width=0.95\linewidth]{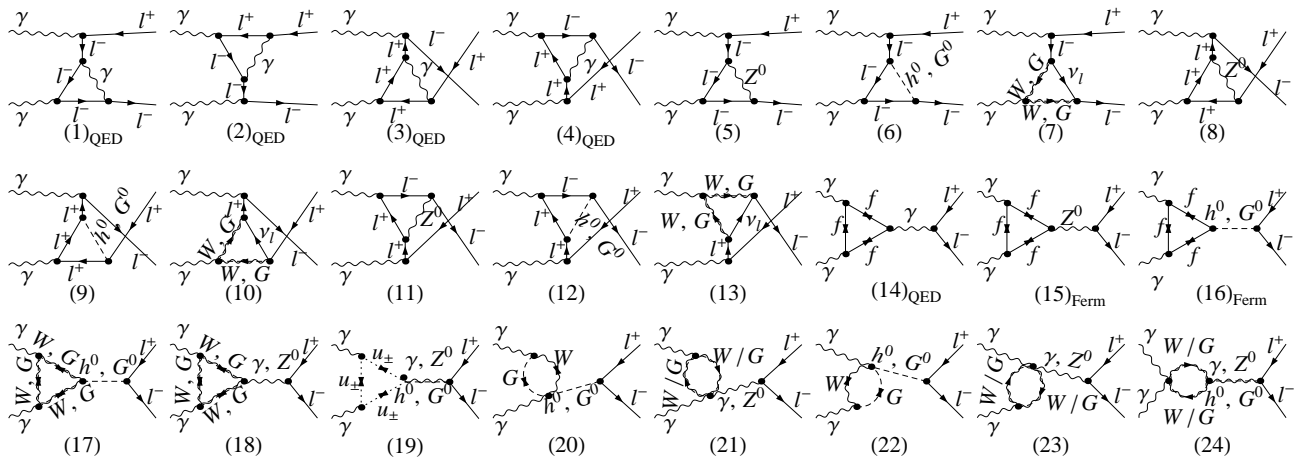}
     \end{center}
     \vspace{-3mm}
\caption{The vertex-correction diagrams contributing to $\gamma \gamma \rightarrow \ell^{-}\ell^{+}$.}\label{fig:diagvert}
\end{figure*}

Here, explicit analytical expressions of the full virtual contributions are not presented as they are very complicated. Instead,  a list of all one-loop Feynman diagrams are provided by the \textsc{FeynArts}. There are 236 one-loop Feynman diagrams in total (12 self-energy+ 26 box-type + 198 vertex-type) as shown in Figs.~\ref{fig:diagself} to~\ref{fig:diagvert}. In diagrams with two arrows on the same lines of the loop, particles are running both counter clock and counterclockwise. The internal lines are labeled as follows: $f$ stands for $\{e, \mu, \tau, u, c, t, d, s, b\}$ and a $\ell$ for leptons $\{\mu,\tau\}$; ($G^0, G$) are neutral/charged Goldstone bosons; $u_{\pm}$ indicates the ghosts. The dashed-lines represent Higgs boson and Goldstone bosons, and the wavy-lines indicate gauge vector bosons ($\gamma$ and $Z$, $W^\pm$). The Mandelstam variables given in~\eqref{eq:manvar} are used. We can also topologically divide one-loop contributions into $\hat{s}$, $\hat{t}$ and $\hat{u}$-channel diagrams with the mediator of gauge bosons ($\gamma$, $Z$, $W^\pm$), neutral Higgs bosons ($h^0$), and neutral/charged Goldstone bosons ($G^0,G^\pm$).

Firstly, we present the self-energy diagrams in Fig.~\ref{fig:diagself}, in which they include all possible loops of leptons, $\gamma$, $Z$, $W^\pm$,  Higgs and Goldstone bosons on the charged lepton propagators. %For the renormalization, only the $\gamma Z$-mixing energy is also needed.
Secondly, in Fig.~\ref{fig:diagbox}, we give the irreducible one-loop diagrams, the so-called box-type contributions. These include all possible loops of leptons, $\gamma$, $Z$, $W^\pm$, Higgs and Goldstone bosons. They are mainly from $\hat{t}$ and $\hat{u}$-channel.

Finally, we give the vertex-correction diagrams in Fig.~\ref{fig:diagvert}. They consist of triangle corrections to $\hat{t}$-channel charged lepton exchange, bubbles and triangle vertices attached to the final state via an intermediate $\gamma$, $Z$, Higgs and Goldstone bosons. We can classify them into three distinct groups. The first type is the vertex corrections $A \ell^{*}\overline{\ell}$ and $A \ell \overline{\ell}^{*}$ in the $\hat{t}$ and $\hat{u}$ channels, where asterisk marks the off-shell field. In Fig.~\ref{fig:diagvert}, the vertex corrections $A \ell^{*}\overline{\ell}$ are given with diagrams $(1),(2)$ and $(5)-(7)$, while the vertex corrections $A \ell \overline{\ell}^{*}$ are shown with diagrams $(3),(4)$ and $(8)-(13)$. The second and third types are the vertex corrections $A A Z^{*}/G^{0*}$ and $A A h^{0*}$, respectively. These are $\hat{s}$-channel contributions shown in diagrams $(14)$-$(24)$ in Fig.~\ref{fig:diagvert}. However, the fermion loop contributions to these vertices are proportional to the final fermion mass and thus can be neglected.

\begin{figure}[hbt]
    \begin{center}
\includegraphics[width=1\linewidth]{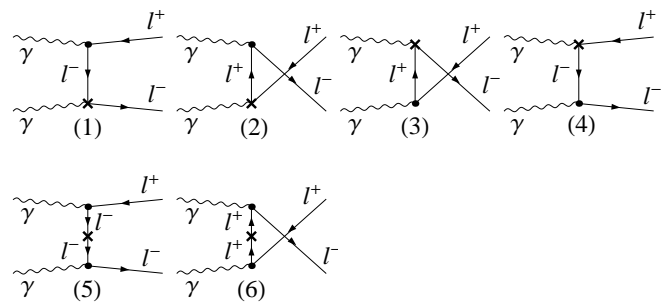}
     \end{center}
     \vspace{-3mm}
\caption{The counterterm diagrams for $\gamma\gamma \rightarrow \ell^{-}\ell^{+}$.}\label{fig:counter}
\end{figure}

The total amplitude for the virtual contributions can be given by the summation of self-energy, box-type and triangle-type contributions, represented as
\begin{equation}\label{eq:totalM}
{\cal \delta M}_{\text{virt}}={\cal M}_{\bigcirc} + {\cal M}_{\Box}+  {\cal M}_{\triangle}.
\end{equation}
The differential cross section for virtual one-loop contributions can be obtained via
\begin{equation} \label{eq:dsigmavirt}
d\hat{\sigma}_{\text{virt}}^{\gamma\gamma\rightarrow \ell^{-}\ell^{+}}=\frac
{1}{16\pi \hat{s}^{2}} \biggl(\frac{1}{4}\biggr) \sum_{hel} 2 \text{Re}\bigl[{\cal M}_{\text{Born}}^*{\cal \delta M}_{\text{virt}}\bigr]d\hat{t}
\end{equation}
where $|{\cal \delta M}_{\text{virt}}|^2$ is not included since it is very small. The one-loop Feynman diagrams, which form the virtual $\mathcal{O}(\alpha)$ corrections ${\cal \delta M}_{\text{virt}}$, have been computed in 't Hooft-Feynman gauge using the on-shell renormalization scheme described in Ref.~\cite{Denner93}. The virtual corrections contain both infrared (IR) and ultraviolet (UV) divergences. The UV divergences are treated via dimensional regularization~\cite{Hooft72}. The counter-terms are taken as in diagrams of Fig.~\ref{fig:counter}.
As described in Ref.~\cite{Denner93}, we have used all Feynman rules including the counter-terms as well as the renormalization conditions in this work. The fields and parameters redefinition is performed in the on-shell scheme. This redefinition transforms the Lagrangian into a bare and counter-term. After the renormalization procedure, we get the virtual part as UV-finite. This can be checked both analytically and numerically. Though, there still is the soft IR singularity, originated from virtual photonic loop correction. The IR singularity is regulated by adding a photon mass parameter, $m_\gamma$. From the Kinoshita-Lee-Nauenberg theorem~\cite{Kinoshita62,Lee64}\footnote{It is also well-known that this was demonstrated perturbatively in QED by Schwinger~\cite{Schwinger49}.}, it is cancelled in the limit $m_\gamma \rightarrow 0$ by adding corrections of the real photon bremsstrahlung.

\subsubsection{Real corrections}
Real photon radiation in $\gamma\gamma \rightarrow \ell^{+}\ell^{-}$ leads to the kinematically
different process which can be expressed as
\begin{equation} \label{eq:gammagammaHHgamma}
\gamma(p_1,\lambda_1)\gamma(p_2,\lambda_2)\rightarrow\ell^{+}(k_1,\sigma_1) \ell^{-}(k_2,\sigma_2) \gamma(k_3,\lambda_3),
\end{equation}
where $k_3$ denotes the radiated photon four-momenta.
\begin{figure}[ht]
    \begin{center}
\includegraphics[width=1\linewidth]{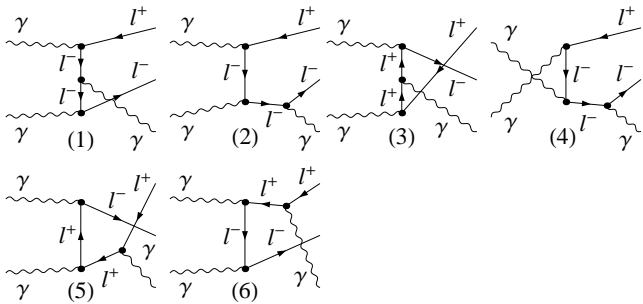}
     \end{center}
     \vspace{-3mm}
\caption{The Feynman diagrams for the real photon radiation.}\label{fig:radiation}
\end{figure}
Figure~\ref{fig:radiation} shows the relevant diagrams. The lowest-order cross section of real-photon radiation process~\eqref{eq:gammagammaHHgamma} yields an $\mathcal{O}(\alpha)$ correction to process $\gamma\gamma \rightarrow \ell^{+}\ell^{-}$. The differential cross section is given by
\begin{equation} \label{eq:difsigma}
\begin{split}
\biggl(\frac{d\hat{\sigma}}{d\Omega_3 }\biggr)_{\text{real}}^{\gamma\gamma\rightarrow \ell^{-}\ell^{+}\gamma} = &\frac
{1}{2 \hat{s}} \sum_{\lambda_{1,2},\sigma_{1,2},\lambda_3}\frac{1}{4} (1+P_1 \lambda_1)(1+P_2 \lambda_2)\\
& \times |{\cal M}_{\gamma\gamma \rightarrow \ell^{+}\ell^{-}\gamma}^{\lambda_1,\lambda_2,\sigma_1,\sigma_2,\lambda_3}|^2
\end{split}
\end{equation}
with the 3-particle phase-space integral is
\begin{equation} \label{eq:domega3}
\begin{split}
\int d\Omega_3 = \prod_{i=1}^{3} \int \frac{d^3 \vec{k}_i}{(2 \pi)^3 2k_i^0 }(2 \pi)^4
\delta\biggl(p_1+p_2-\sum_{j=1}^3 k_j \biggr).
\end{split}
\end{equation}

According to the radiated photon energy $k_3^0=\sqrt{|\overrightarrow{k_3}|^2+m_\gamma^2}$, we can divide the bremsstrahlung phase space into soft and hard regions. Correction from the real photon radiation then reads
\begin{equation} \label{eq:dsigmaSB}
d\hat{\sigma}_{\text{real}}^{\gamma\gamma\rightarrow \ell^{-}\ell^{+}\gamma}=d\hat{\sigma}_{\text{soft}}(\Delta_s)+d\hat{\sigma}_{\text{hard}}(\Delta_s)
\end{equation}
where $\Delta_s$ denotes the soft cut-off energy parameter $\Delta_s=\Delta E_\gamma/(\sqrt{\hat{s}}/2)$. The radiated photon is called soft when $k_3^0<\Delta E_\gamma=\Delta_s \sqrt{\hat{s}}/2 $, while it is hard if $k_3^0>\Delta E_\gamma$. The approximation formula~\cite{Hooft79,Denner93}
\begin{equation} \label{eq:dsoft}
\begin{split}
d\hat{\sigma}_{\text{soft}}=-d\hat{\sigma}_{\text{Born}}\frac{\alpha Q^2_\ell}{2\pi^2} \int_{|\overrightarrow{k_3}|\leq \Delta E_\gamma} \frac{d^3k_3}{2k^0_3} \biggl[\frac{k_1}{k_1\cdot k_3}-\frac{k_2}{k_2\cdot k_3}\biggl]^2
\end{split}
\end{equation}
gives the soft photon correction, where $d\hat{\sigma}_{\text{Born}}$ denotes the Born-level differential cross section and $\Delta E_\gamma$ satisfies $k^0_3\leq\Delta E_\gamma \ll\sqrt{\hat{s}}$. Integrating the soft photon phase space in the center-of-mass system yields
\begin{equation} \label{eq:dsoft0}
\begin{split}
d\hat{\sigma}_{\text{soft}}=\delta_{\text{soft}}d\hat{\sigma}_{\text{Born}}
\end{split}
\end{equation}
with
\begin{equation} \label{eq:dsoft1}
\begin{split}
\delta_{\text{soft}}=&-\frac{\alpha}{\pi}Q^2_\ell\biggl[2 \ln\biggl( \frac{2\Delta E_\gamma}{m_\gamma}\biggr) \biggl(1+\ln\biggl( \frac{m^2_\ell}{\hat{s}}\biggr)\biggr)\\
&+ \frac{1}{2}\ln^2\biggl( \frac{m^2_\ell}{\hat{s}}\biggr)+\ln\biggl( \frac{m^2_\ell}{\hat{s}}\biggr)+ \frac{\pi^2}{3}
\biggr].
\end{split}
\end{equation}

The real corrections are independent of the soft cut-off parameter $\Delta_s$ despite both the soft and hard photon parts depend on this parameter. On the other hand, adding the virtual and soft contributions cancels the IR regulator $m_\gamma$ dependence. The results now depend on $\Delta_s$ (i.e., $\Delta E_\gamma$), so that contribution of the hard photon radiation must also be included for dropping out this dependency.

\subsubsection{Classification of full corrections}
We can separate the UV and IR finite result for $\mathcal{O}(\alpha)$ corrections into three parts:
\begin{equation} \label{eq:dsigmaNLO}
\begin{split}
d\hat{\sigma}_{\text{NLO}}^{\gamma\gamma\rightarrow \ell^{+}\ell^{-}}&=d\hat{\sigma}_{\text{virt}}(m_\gamma)+d\hat{\sigma}_{\text{soft}}(m_\gamma,\Delta_s)\\
&+d\hat{\sigma}_{\text{hard}}(\Delta_s).
\end{split}
\end{equation}
The first is from virtual (loop) contribution, the second from the soft photon emission, and the third from the real hard photon Bremsstrahlung. This form depends neither on the IR regulator $m_\gamma$ nor on the soft cut-off parameter $\Delta_s$.

\begin{figure*}[hbt]
    \begin{center}
\includegraphics[scale=0.42]{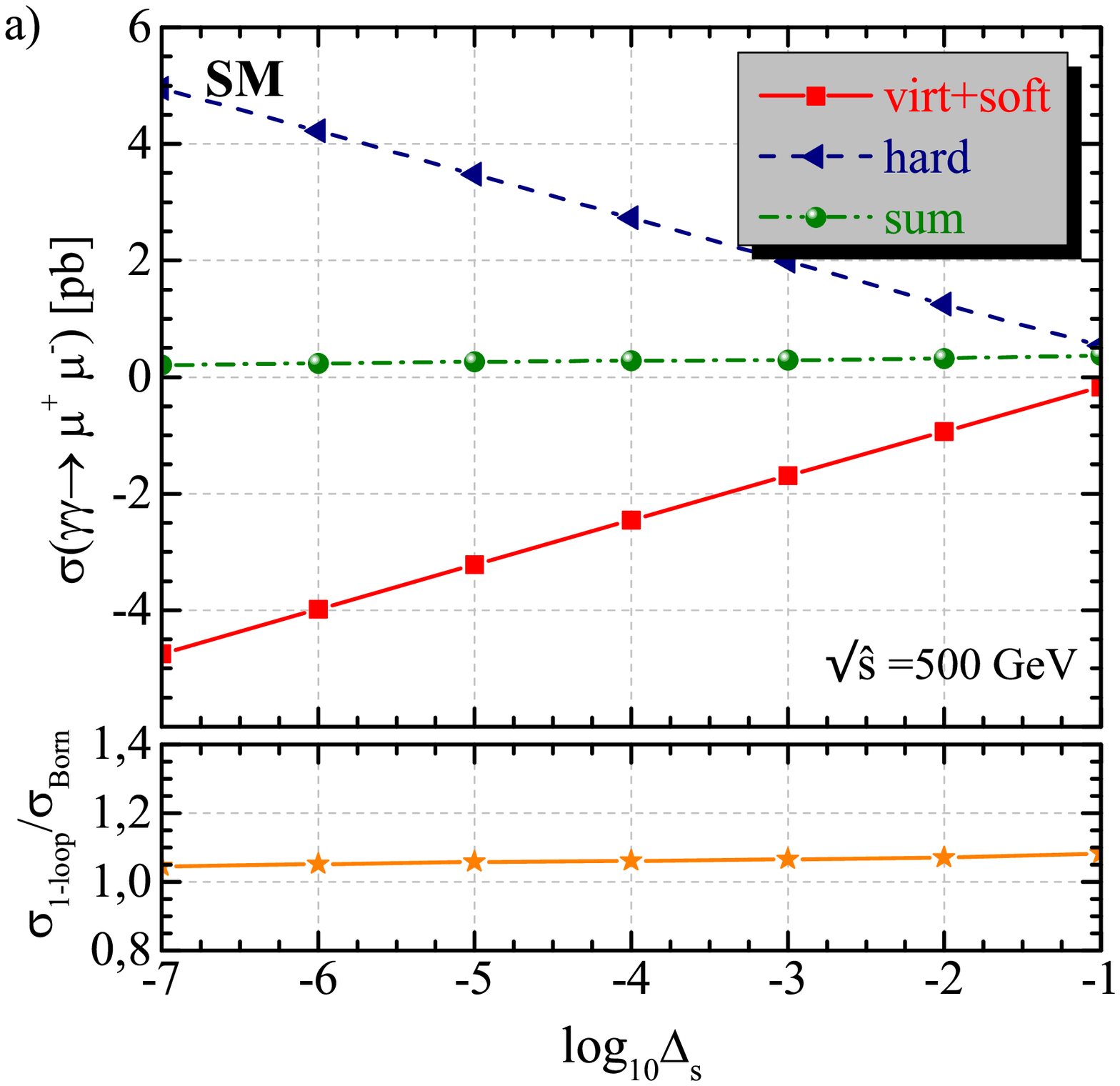}
\includegraphics[scale=0.42]{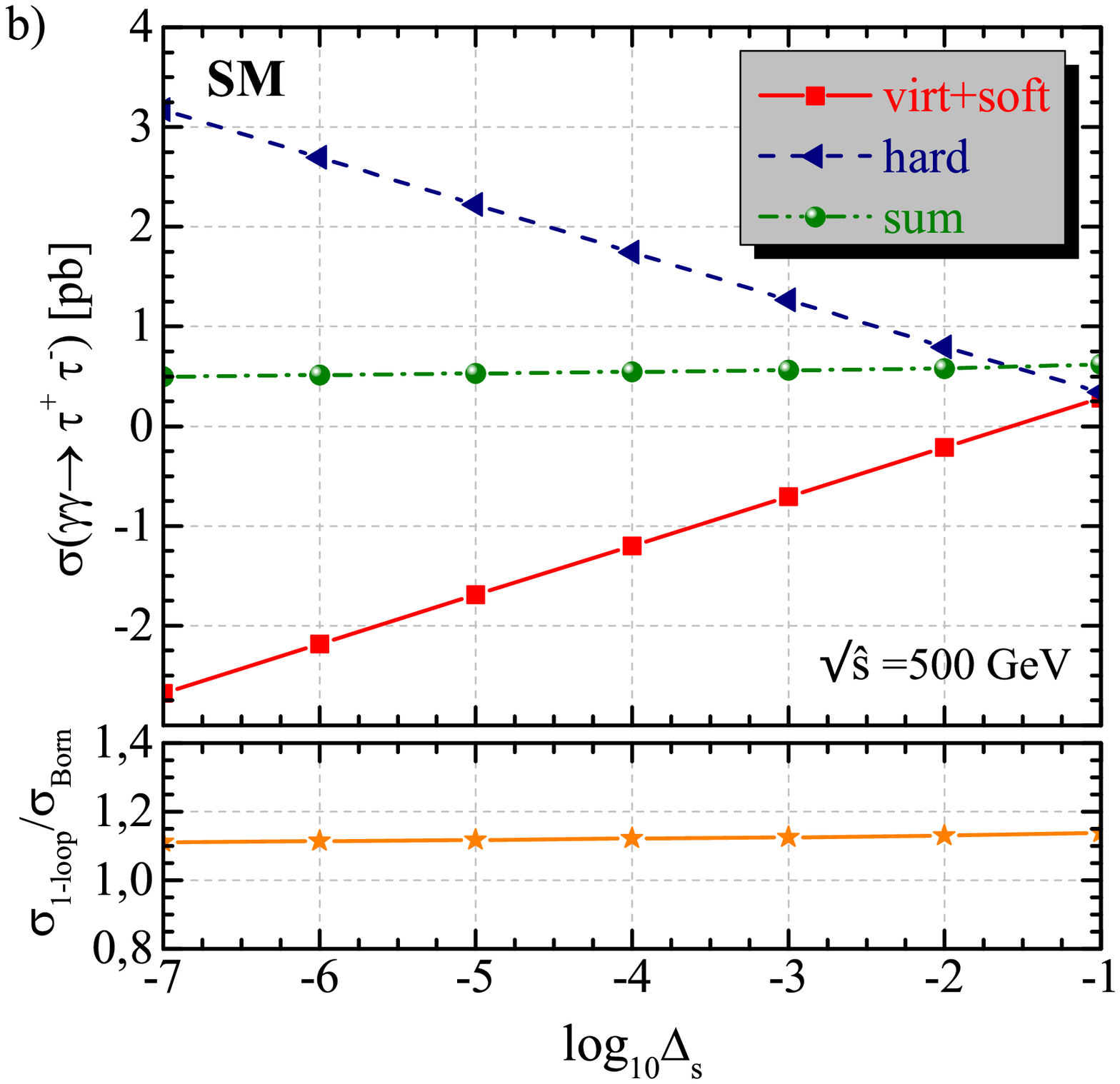}
     \end{center}
     \vspace{-4mm}
\caption{The virtual plus soft and hard photon radiation corrections to processes a) $\gamma\gamma \to \mu^- \mu^+  $ and b) $\gamma\gamma \to \tau^- \tau^+  $ as a function of $\Delta_s$ in the $\alpha(0)$-scheme.}
\label{fig:delts}
\end{figure*}
We have numerically inspected that our results are independent of $m_\gamma$ or on $\Delta E_\gamma=\Delta_s \sqrt{\hat{s}}/2 $. In Fig.~\ref{fig:delts}, we present the virtual plus soft photon, the hard photon radiation and the total one-loop corrections as a function of $\Delta_s$ at $\sqrt{\hat{s}}=500\gev$ for both $\gamma\gamma \to \mu^- \mu^+  $ and $\gamma\gamma \to \tau^- \tau^+ $  processes. We can see from these figures that although the virtual plus soft correction and the hard photon radiation correction strongly depend on $\Delta_s$, the total correction is independent of this parameter for both processes. Moreover, the relative one-loop correction $\hat{\sigma}_{\text{1-loop}}/\hat{\sigma}_{\text{Born}}$ is also stable around $1.06$ and $1.12$ for $\gamma\gamma \to \mu^- \mu^+  $ and $\gamma\gamma \to \tau^- \tau^+ $, respectively.

The Feynman diagrams of the virtual contributions can also be classified according to gauge-invariant subsets: QED-corrections $\delta_{\text{QED}}$ and Weak corrections $\delta_{\text{Weak}}$. Thus, the total virtual correction becomes
\begin{equation} \label{eq:dtotal}
\delta_{\text{Total}}=\delta_{\text{QED}}+\delta_{\text{Weak}}.
\end{equation}
The process $\gamma\gamma \rightarrow  \ell^+ \ell^-$ can be treated in pure QED. The QED corrections include virtual-photon exchange, real-photon emission, and the corresponding counterterms. All QED-like diagrams, that is, those with only $A$ and $l$ fields as virtual lines, form a subset of gauge-invariant. We consider the one-loop QED corrections $\delta_{\text{QED}}$  as the sum of the soft-photon contribution and the contribution of diagrams $(1)$ and $(5)$ of Fig.~\ref{fig:diagself}, $(1)$ and $(5)$ of Fig.~\ref{fig:diagbox} and $(1)-(4)$ of Fig.~\ref{fig:diagvert}. The remaining contributions (non-QED corrections) are defined as weak corrections $\delta_{\text{Weak}}$ that involve the massive $Z^0$ and $W^\pm$ gauge bosons. They can be further divided into two subsets as follows:
\begin{equation} \label{eq:dweak}
\delta_{\text{Weak}}=\delta_{\text{ferm}}+\delta_{\text{bos}}
\end{equation}
where $\delta_{\text{ferm}}$ denotes fermionic contributions and $\delta_{\text{bos}}$ is bosonic contributions. This kind of decomposition was previously made for the $\gamma\gamma \to t \bar{t} $ process in Ref.~\cite{Denner96}.  Because the number of fermion generations is a free parameter in the SM, each generation provides a subset with gauge-invariant of one-loop diagrams created with fermionic loops. This correction come from diagrams $(15)$ and $(16)$ of Fig.~\ref{fig:diagvert}. Finally, the all remaining, except those described above, also form a gauge-invariant subset termed bosonic corrections $\delta_{\text{bos}}$.

As a result, the factor $\delta_{\text{Total}}$ represents the full relative $\mathcal{O}(\alpha)$ correction. The one-loop corrected cross-section can be factorized into the Born-level cross section and the relative correction. Thus, the integrated cross-section $\sigma_{\text{1-loop}}$ at one-loop level can be separately defined by
\begin{equation} \label{eq:sigma}
\begin{split}
\sigma_{\text{1-loop}}&=\sigma_{\text{Born}}(1+\delta_{\text{Total}})\\
&=\sigma_{\text{Born}}(1+ \delta_{\text{QED}}+\delta_{\text{Weak}})\\
&=\sigma_{\text{Born}}(1+\delta_{\text{QED}}+\delta_{\text{ferm}}
+\delta_{\text{bos}}),
\end{split}
\end{equation}
leading to
\begin{equation} \label{eq:dtotal}
\begin{split}
\delta_{\text{Type}}&=\frac{\sigma^{\text{Type}}_{\text{1-loop}}-\sigma_{\text{Born}}}{\sigma_{\text{Born}}}
\end{split}
\end{equation}
where $\text{Type}$ can be ``Total'', ``QED'', ``Weak'', ``ferm'', ``bos''.

The full EW $\mathcal{O}(\alpha)$ corrections depend on the electromagnetic coupling $\alpha = e^2/(4\pi)$. Different input-parameter schemes can be chosen for $\alpha$ that will affect its value. This can be set in two ways: the fine-structure constant $\alpha(0)$ in the Thompson limit and the running electromagnetic coupling $\alpha(Q^2)$ at a high-energy scale $Q$. As an example, it is possible to use the value of $\alpha(M^2_Z)$ which is calculated by analyzing the experimental ratio $R=\sigma(e^-e^+\rightarrow \text{hadrons})/\sigma(e^-e^+\rightarrow \mu^-\mu^+)$~\cite{Burkhardt95,Eidelman95}. These ways are referred to as $\alpha(0)$-scheme and $\alpha(M^2_Z)$-scheme, respectively. Another choice for $\alpha$ is obtained from the Fermi constant $G_\mu$ as follows
\begin{equation}
\alpha(G_\mu) =\frac{\sqrt{2} G_\mu M^2_W}{\pi} \biggl(1-\frac{M^2_W}{M^2_Z}\biggr),
\end{equation}
which is called a $G_\mu$-scheme. The distinctions between these schemes will be more obvious in the discussion of the relevant $\mathcal{O}(\alpha)$ corrections.

\subsection{Parent Process $e^+e^-\rightarrow \gamma\gamma\rightarrow  \ell^+ \ell^-$}
The $\gamma\gamma$ collision would be one of the interesting processes in the future colliders with TeV energy scale. When a polarized laser beam undergoes Compton-scattering with a polarized electron beam at a facility of this kind, each electron is effectively converted into a high-energy polarized photon from its energy fraction. Both effective luminosity and energy of $\gamma\gamma$ collisions from back-scattered laser beams are expected to be comparable with that of the primary $e^-e^+$ collisions. The high energy luminosity and polarization of backscattered laser beams hence have the potential to make $\gamma\gamma$ collisions a key component of the physics program of the future linear collider. This feature would allow detailed works of a large array, including polarized beams, of high energy $\gamma\gamma$ and $\gamma e$ collisions.

In this context, $\gamma\gamma\to \ell^+ \ell^-$ can be generated as a subprocess of $e^-e^+$ collision. For the parent process $e^+e^-\rightarrow\gamma\gamma\rightarrow \ell^+ \ell^-$, the total cross-section, folding $\hat{\sigma}(\gamma\gamma\rightarrow \ell^+ \ell^-)$ with the photon luminosity
\begin{equation}
\frac{dL_{\gamma\gamma}}{dz}=2z\int_{z^2/x_{max}}^{x_{max}}\frac{dx}{x}F_{\gamma/e}(x)F_{\gamma/e}\left(\frac{z^2}{x}\right)
,
\end{equation}
can be written as
\begin{align} \label{eq:total_cross}
\begin{split}
&\sigma^{e^+e^-\rightarrow\gamma\gamma\rightarrow \ell^+ \ell^-}(s)=\\
&\int_{(2m_{\ell^\pm})/\sqrt{s}}^{x_{max}} dz \frac{dL_{\gamma\gamma}}{dz}~ \hat{\sigma}( \gamma\gamma\rightarrow  \ell^+ \ell^-;\; \hat{s}=z^2s ),
\end{split}
\end{align}
where $F_{\gamma/e}(x)$ denotes the photon structure function. This entity is represented for the initial unpolarized electrons and laser photon beams by the most promising Compton backscattering as \cite{Ginzburg83,Ginzburg84,Telnov}
\begin{equation}
\begin{split}
F_{\gamma/e}(x)=&\frac{1}{D(\zeta)}\biggl[1-x+\frac{1}{1-x}-
\frac{4x}{\zeta(1-x)}\\
&+\frac{4x^{2}}{\zeta^{2}(1-x)^2}\biggr],
\end{split}
\end{equation}
with
\begin{equation}
D(\zeta)=(1-\frac{4}{\zeta}-\frac{8}{{\zeta}^2})\ln{(1+\zeta)}+\frac{1}{2}+
  \frac{8}{\zeta}-\frac{1}{2{(1+\zeta)}^2},
\end{equation}
where $\zeta=\frac{2\sqrt{s} \omega_0}{m_e^2}$, $\omega_0$ denotes the laser-photon energy and $x$ is the fraction of the energy of the incident electron carried by the backscattered photon. The maximum fraction of energy carried by the backscattered photon is $x_{max}=2 \omega_{max}/\sqrt{s}=\zeta/(1+\zeta)$. We take $\omega_0$ to maximize the backscattered photon energy in our calculations, such that it is not spoiling the luminosity via electron-positron pair creation. This procedure requires ${\zeta}\leq2(1+\sqrt{2})\simeq4.8$, and hence we have $x_{max}\simeq 0.83$, and $D(\zeta)\approx 1.8397$.

\section{Numerical Results And Discussions} \label{sec:results}
We discuss in detail the numerical results of the muon-pair and tau-pair production in photon-photon collisions, by considering complete one-loop corrections, including soft and hard QED radiation. We present the Born-level and one-loop cross sections by considering the pure QED and weak corrections, separately, as a function of the center-of-mass energy. In addition, we complement our calculation by partitioning the weak corrections into the purely fermionic loops together with the fermionic part of the counterterms and the remaining weak bosonic contributions. We show their effect on the total cross section with the relative corrections previously defined in Eq.~\eqref{eq:dtotal}. Angular and polarization distributions of the process are also presented.

We implement the following input parameters for the numerical calculations~\cite{PDG20}:
\begin{equation*}\label{eq:SMpar}
\begin{tabular}{ll}
%\hline
$\alpha(0)= 1/137.03599907$, & $G_F = 1.1663787(6)\times10^{-5}\gev^{-2}$,\\
$M_W= 80.385\gev$, &$m_e=0.510998928\mev$,  \\
$M_Z= 91.1876\gev$, &$m_\mu=105.6583715\mev$, \\
$M_h= 125\gev$, &$m_\tau=1.77682\gev$,\\
$m_u=73.56\mev$,&$m_d=73.56\mev$,  \\
$m_c=1.275\gev$, & $m_s=95\mev$, \\
 $m_t=173.21\gev$, & $m_b=4.66\gev$.\\
%\hline
\end{tabular}
\end{equation*}

Furthermore, we set the soft cutoff parameter as $\Delta_s = 10^{-3}$, and $|\cos\theta| < 0.99$ for the range of scattering angles of the final particles. During our numerical evaluation, we use the $\alpha(0)$-scheme where $\alpha$ is inputted in the Thompson limit, yet we also present a comparison with the results obtained in the $G_\mu$-scheme.

\begin{table}[!ht]
\caption{The triple comparison between \textsc{FeynArts}$\&$\textsc{FormCalc} (FA$\&$FC), \textsc{Whizard} and \textsc{CalcHEP} of the Born-level and hard photon bremsstrahlung cross section calculations.}\label{table:toolscomp}
\centering
\begin{ruledtabular}
\begin{tabular}{llll}
$~~~\sqrt{s}$ & $250\gev$ & $500\gev$ & $1000\gev$\\
\hline
&\multicolumn{3}{c}{$\sigma^{\text{Born}} (\gamma\gamma \to \mu^- \mu^+)$ [pb]}\\
%\hline
FA$\&$FC &17.940(4)  & 4.4851(3) &1.1212(8) \\
\textsc{Whizard}  &17.940(6)  & 4.4850(8) &1.1212(6)\\
\textsc{CalcHEP}  &17.940     & 4.4851    &1.1213 \\
&\multicolumn{3}{c}{$\sigma^{\text{Born}} (\gamma\gamma \to \tau^- \tau^+)$ [pb]}\\
%\hline
FA$\&$FC &17.903(7) & 4.4828(3) &1.1211(4) \\
\textsc{Whizard}  &17.903(8) & 4.4830(1) &1.1211(3)\\
\textsc{CalcHEP}  &17.904    & 4.4828    &1.1211  \\
\hline
&\multicolumn{3}{c}{$\sigma^{\text{hard}-\gamma} (\gamma\gamma \to \mu^- \mu^+\gamma)$ [pb]}\\
%\hline
FA$\&$FC &7.2448(1)  & 1.9892(5) &0.54168(9) \\
\textsc{Whizard}  &7.2444(1)  & 1.9874(1) &0.54194(8)\\
\textsc{CalcHEP}  &7.2432     & 1.9894    &0.54194 \\
&\multicolumn{3}{c}{$\sigma^{\text{hard}-\gamma} (\gamma\gamma \to \tau^- \tau^+\gamma)$ [pb]}\\
%\hline
FA$\&$FC & 4.3588(7) & 1.2664(9) &0.36082(8) \\
\textsc{Whizard}  & 4.3560(6) & 1.2657(0) &0.36059(6)\\
\textsc{CalcHEP}  & 4.3517    & 1.2656    &0.36417  \\
%\hline
\end{tabular}
\end{ruledtabular}
\end{table}
\begin{figure*}[!hbt]
    \begin{center}
\includegraphics[scale=0.43]{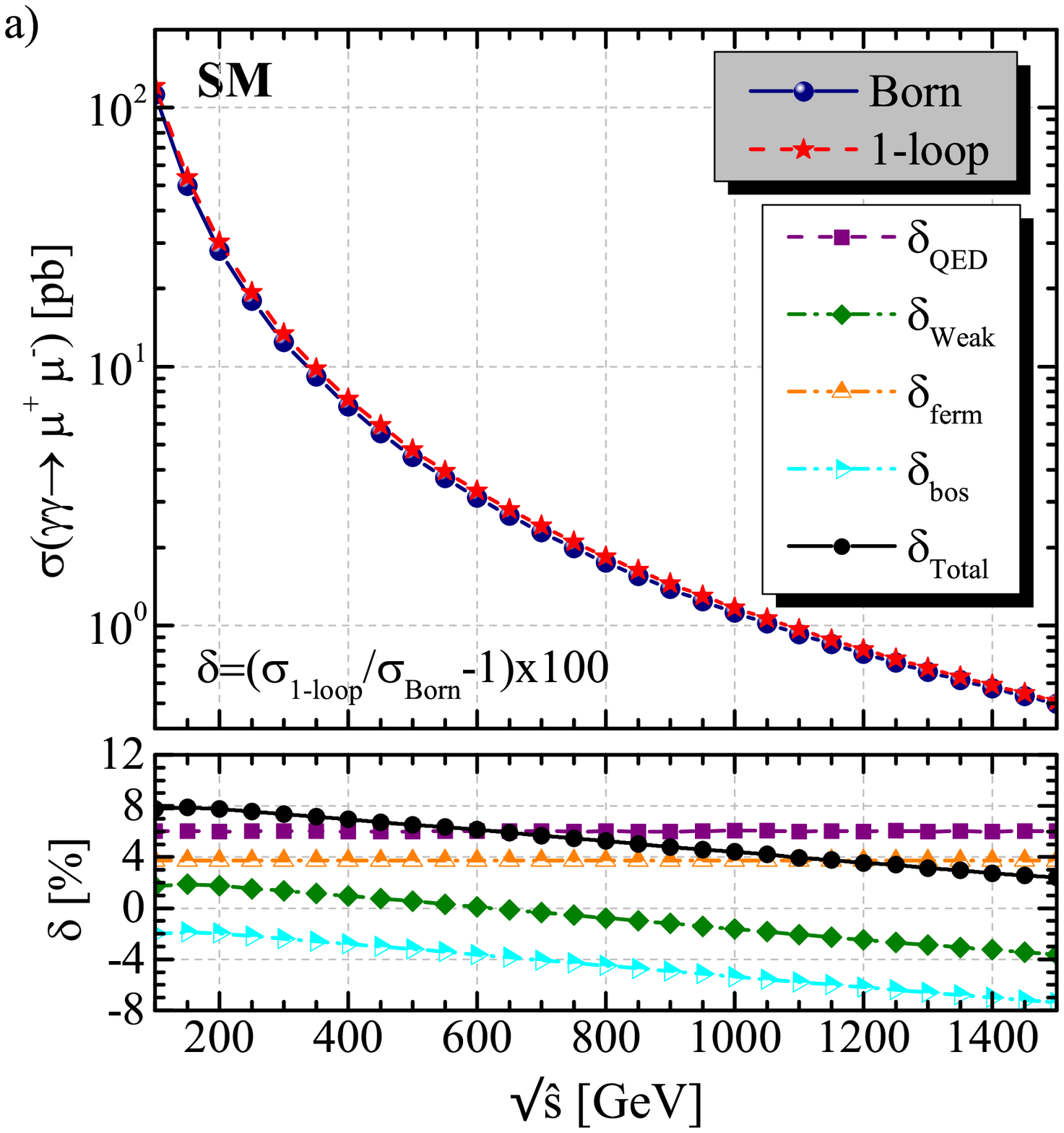}
\includegraphics[scale=0.43]{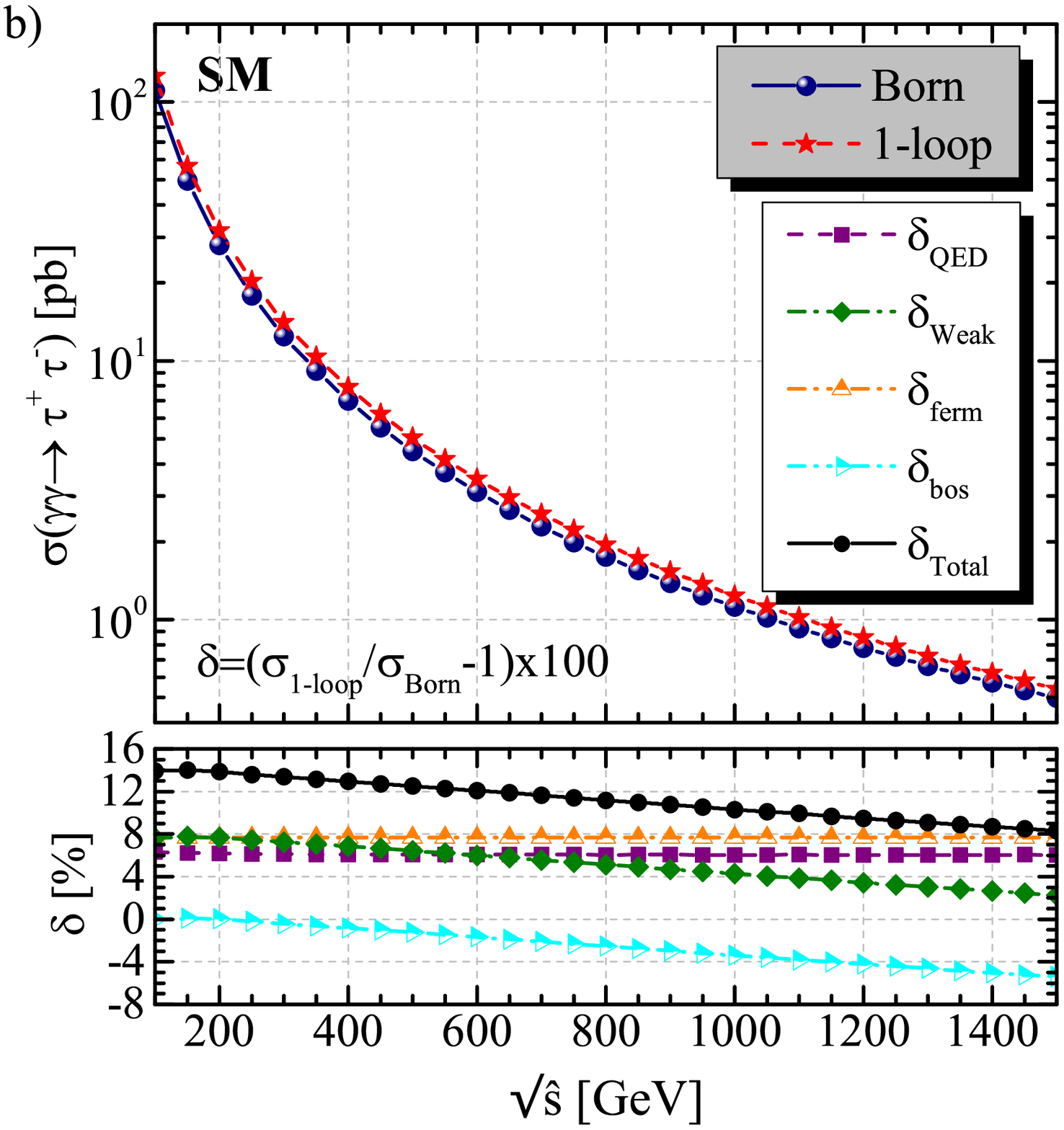}
     \end{center}
     \vspace{-4mm}
\caption{(color online). Born-level and one-loop cross sections as a function of center of mass energy for (a) $ \gamma \gamma \to \mu^- \mu^+$ and (b) $ \gamma \gamma \to \tau^- \tau^+$. We also show relative corrections in a percentage at the bottom panel.}
\label{fig:cs}
\end{figure*}
\begin{figure*}[htb]
    \begin{center}
\includegraphics[scale=0.43]{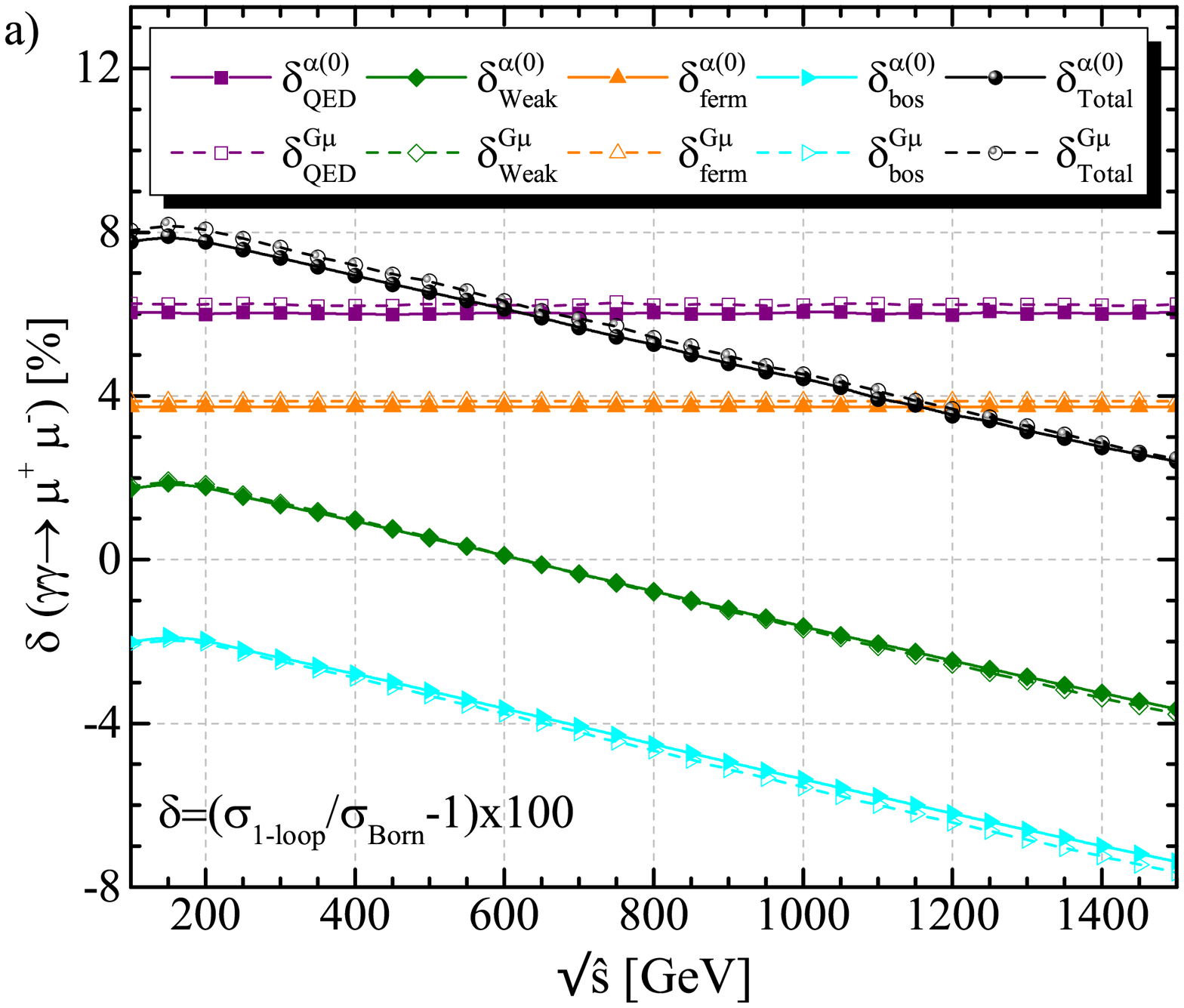}
\includegraphics[scale=0.43]{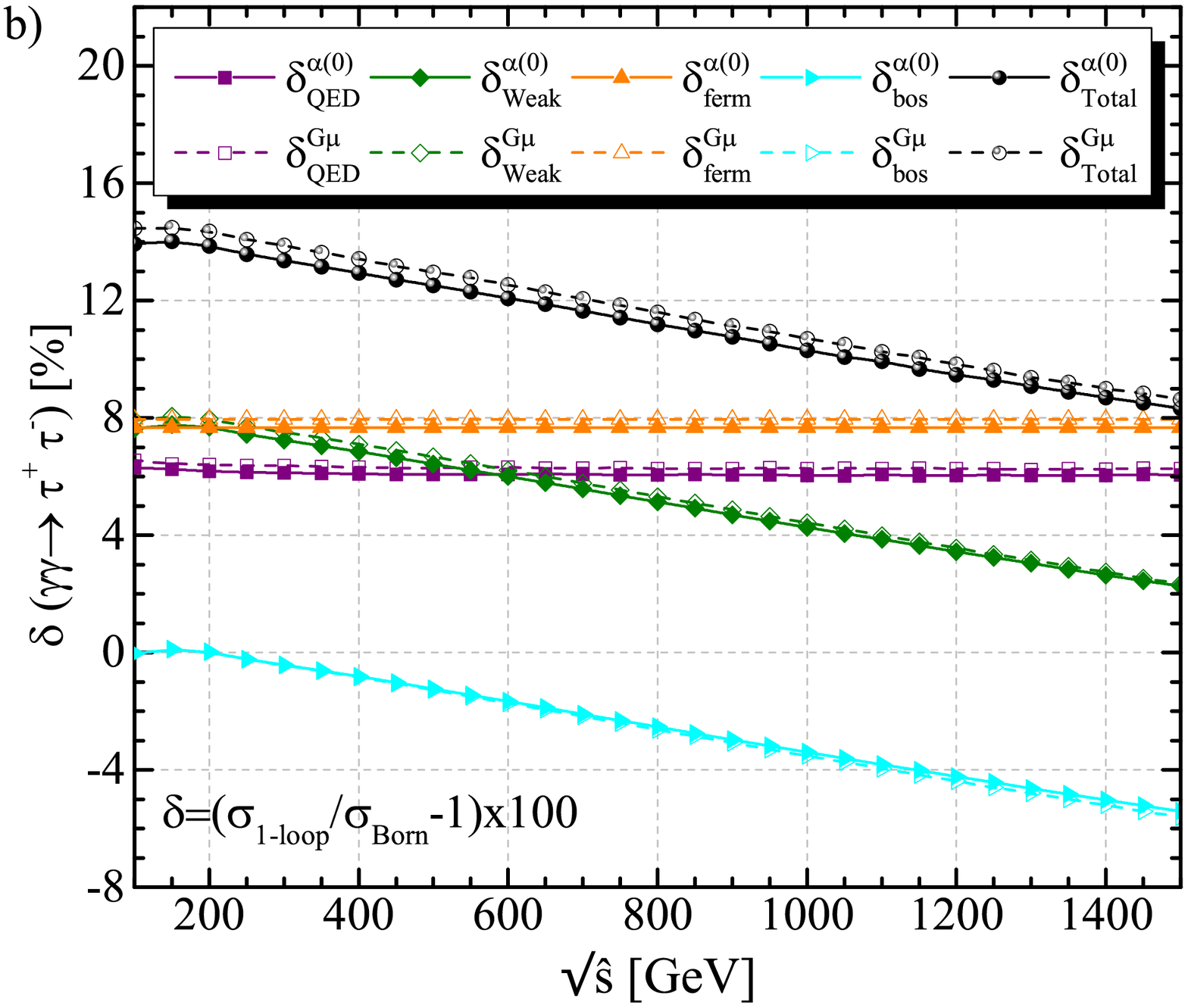}
     \end{center}
     \vspace{-4mm}
\caption{(color online). Relative corrections in the $\alpha(0)$-scheme and $G_\mu$-scheme as a function of center of mass energy for (a) $ \gamma \gamma \to \mu^- \mu^+$ and (b) $ \gamma \gamma \to \tau^- \tau^+$}
\label{fig:deltaschemes}
\end{figure*}
First of all, we compare the numerical results of FeynArts$\&$FormCalc with the ones obtained by using CalcHEP and WHIZARD for the cross-section of Born-level and hard photon bremsstrahlung. In Table~\Ref{table:toolscomp}, we present the triple comparison for Born level process $\gamma\gamma \to \ell^- \ell^+$ and hard process $\gamma\gamma \to \ell^- \ell^+ \gamma$ at $\sqrt{\hat{s}}=250\gev$, $500\gev$ and $1000\gev$. We have found an excellent agreement for the Born-level cross section with the aforementioned packages. We have obtained very good agreement within four to five digits for hard photon bremsstrahlung cross-section.

In Figs.~\ref{fig:cs}(a)-(b), we present the Born-level and one-loop cross sections for processes $ \gamma \gamma \to \mu^- \mu^+$ and $ \gamma \gamma \to \tau^- \tau^+$ as a function of center of mass energy, respectively. Moreover, to describe each different type of correction in the total cross-section quantitatively, we plot the corresponding relative corrections in the same figures. Since the center of mass energy starts at a greater value than the threshold, all curves start from their maximum values and decrease with the increment of $\sqrt{\hat{s}}$. As the $\sqrt{\hat{s}}$ goes from 100 GeV to 1.5 TeV, one-loop cross section decreases from 120.83 pb to 0.51 pb for $\gamma \gamma \to \mu^- \mu^+$ and 126.16 pb to 0.54 pb for $ \gamma \gamma \to \tau^- \tau^+$. The full EW corrections give a positive contribution to the total cross section in the parameter regions. These results show that the production rate of $\gamma\gamma \rightarrow \ell^{-}\ell^{+}$ is larger by one order of magnitude than from the $e^- e^+$-collision mode (see, Refs.~\cite{Bondarenko18,Bondarenko20}).

The fermionic corrections are around $+3.7\%$ for $\gamma \gamma \to \mu^- \mu^+$ and $+7.7\%$ for $ \gamma \gamma \to \tau^- \tau^+$, and depend on the center of mass energy weakly. Whereas the bosonic corrections are about $-2\% (-0.01\%)$ close to threshold, they decrease rapidly with the increasing center of mass energies, eventually reaching about $-7.34\% (-5.41\%)$ at an energy of 1.5 TeV for $\gamma \gamma \to \mu^- \mu^+$ ($ \gamma \gamma \to \tau^- \tau^+$). Consequently, the fermionic and bosonic corrections are partially cancelled, as they are combined to form the full weak corrections. The fermionic corrections stay almost stable even for high energies. The bosonic corrections supply a determinative contribution to weak corrections at high energies. The QED corrections are about $+6\% $ at the first point, which hardly changes up to 1500 GeV for both processes. The QED corrections independent of the center of mass energy for $s\gg m^2_\ell$, because all mass singularities cancel, and the only scale that survives is $s$. Both QED and weak corrections are partially compensate each other in the EW corrections, yielding relative corrections of around $+7.77\%$ and $+13.94\%$ at the first point, and $+2.40\%$ and $+8.31\%$ for $\gamma \gamma \to \mu^- \mu^+$ and $ \gamma \gamma \to \tau^- \tau^+$, respectively, at 1500 GeV. Our results show that the weak $\mathcal{O}(\alpha)$ corrections to $\gamma \gamma \to \ell^- \ell^+$ are required to match a percent level accuracy.

In Figs.~\ref{fig:deltaschemes}(a)-(b), we give the relative corrections of processes $ \gamma \gamma \to \mu^- \mu^+$ and $ \gamma \gamma \to \tau^- \tau^+$ for two different schemes, $\alpha(0)$-scheme and $G_\mu$-scheme, as a function of center of mass energy, respectively. The marks$\&$lines corresponding to different relative corrections and schemes are labeled as follows: for $\alpha(0)$-scheme, $\delta_{\text{QED}}^{\alpha(0)}$ (purple square on line), $\delta_{\text{Weak}}^{\alpha(0)}$ (green diamond on line),  $\delta_{\text{ferm}}^{\alpha(0)}$  (orange up-triangle on line), $\delta_{\text{ferm}}^{\alpha(0)}$  (cyan right-triangle on line), and $\delta_{\text{Total}}^{\alpha(0)}$  (black sphere on line). Meanwhile, the calculations in the $G_\mu$-scheme are represented by the same symbols on dashed-lines but with a hollow. As a general comment, it can be emphasized that the relative corrections are almost independent of the choice of scheme, changing only a few percent. The Born and one-loop cross sections increase by up to about $7.4\%$ and $7.8\%$, respectively, in $G_\mu$-scheme as compared to the $\alpha(0)$-scheme. The relative corrections increase by up to about $4\%$, in $G_\mu$-scheme as compared to the $\alpha(0)$-scheme. These results are almost the same for both processes considered in this study.

In Tables~\ref{table:deltamu} and~\ref{table:deltatau}, we list numerical values for the unpolarized cross sections and the relative corrections, calculated in the $\alpha(0)$ and $G_\mu$-schemes for $ \gamma \gamma \to \mu^- \mu^+$ and $ \gamma \gamma \to \tau^- \tau^+$, respectively.
\begin{table}[!ht]
\caption{Born-level, one-loop cross sections (in pb) and relative corrections (in $\%$) of $\gamma \gamma \to \mu^- \mu^+$ in the $\alpha(0)$-scheme and $G_\mu$-scheme for $\sqrt{\hat{s}}=250\gev$, $500\gev$ and $1000\gev$.}\label{table:deltamu}
\centering
\begin{ruledtabular}
\begin{tabular}{crrr}
%\hline
$\sqrt{\hat{s}}$ & $250\gev$ & $500\gev$ & $1000\gev$\\
\hline
&\multicolumn{3}{c}{$\sigma(\gamma\gamma \to \mu^- \mu^+)$ [pb]}\\
%\hline
$\sigma_{\text{Born}}^{\alpha(0)}$   &17.94  &4.49 &1.12 \\
$\sigma_{\text{Born}}^{G_\mu}$       &19.27  &4.82 &1.20\\
$\sigma_{\text{1-loop}}^{\alpha(0)}$ &19.30  &4.78 &1.17 \\
$\sigma_{\text{1-loop}}^{G_\mu}$     &20.78  &5.15 &1.26\\
%\hline
&\multicolumn{3}{c}{$\delta(\gamma\gamma \to \mu^- \mu^+)$ [$\%$]}\\
$\delta_{\text{QED}}^{\alpha(0)}$   &+6.04  &+6.00 &+6.06 \\
$\delta_{\text{QED}}^{G_\mu}$       &+6.26  &+6.26 &+6.22\\
$\delta_{\text{ferm}}^{\alpha(0)}$  &+3.73  &+3.73 &+3.73 \\
$\delta_{\text{ferm}}^{G_\mu}$      &+3.87  &+3.87 &+3.87\\
$\delta_{\text{bos}}^{\alpha(0)}$   &$-$2.20 &$-$3.20 &$-$5.37 \\
$\delta_{\text{bos}}^{G_\mu}$       &$-$2.28 &$-$3.32 &$-$5.56\\
$\delta_{\text{Weak}}^{\alpha(0)}$  &+1.53  &+0.53 &$-$1.63 \\
$\delta_{\text{Weak}}^{G_\mu}$      &+1.59  &+0.55 &$-$1.69\\
$\delta_{\text{Total}}^{\alpha(0)}$ &+7.57  &+6.53 &+4.43 \\
$\delta_{\text{Total}}^{G_\mu}$     &+7.85  &+6.80 &+4.53\\
\end{tabular}
\end{ruledtabular}
\end{table}
\begin{table}[!ht]
\caption{Same as Table~\ref{table:deltamu} but for $\gamma \gamma \to \tau^- \tau^+$.}\label{table:deltatau}
\centering
\begin{ruledtabular}
\begin{tabular}{crrr}
%\hline
$\sqrt{\hat{s}}$ & $250\gev$ & $500\gev$ & $1000\gev$\\
\hline
&\multicolumn{3}{c}{$\sigma(\gamma\gamma \to \tau^- \tau^+)$ [pb]}\\
%\hline
$\sigma_{\text{Born}}^{\alpha(0)}$   &17.90  &4.48 &1.12 \\
$\sigma_{\text{Born}}^{G_\mu}$       &19.23  &4.81 &1.20\\
$\sigma_{\text{1-loop}}^{\alpha(0)}$ &20.33  &5.04 &1.24 \\
$\sigma_{\text{1-loop}}^{G_\mu}$     &21.93  &5.44 &1.33\\
&\multicolumn{3}{c}{$\delta(\gamma\gamma \to \tau^- \tau^+)$ [$\%$]}\\
$\delta_{\text{QED}}^{\alpha(0)}$   &+6.14  &+6.07 &+6.04 \\
$\delta_{\text{QED}}^{G_\mu}$       &+6.37  &+6.28 &+6.27\\
$\delta_{\text{ferm}}^{\alpha(0)}$  &+7.67  &+7.67 &+7.67 \\
$\delta_{\text{ferm}}^{G_\mu}$      &+7.95  &+7.95 &+7.95\\
$\delta_{\text{bos}}^{\alpha(0)}$   &$-$0.23  &$-$1.24 &$-$3.40 \\
$\delta_{\text{bos}}^{G_\mu}$       &$-$0.24  &$-$1.28 &$-$3.52\\
$\delta_{\text{Weak}}^{\alpha(0)}$  &+7.44  &+6.43 &+4.27 \\
$\delta_{\text{Weak}}^{G_\mu}$      &+7.71  &+6.67 &+4.43\\
$\delta_{\text{Total}}^{\alpha(0)}$ &+13.57 &+12.51&+10.31 \\
$\delta_{\text{Total}}^{G_\mu}$     &+14.08 &+12.95&+10.69\\
%\hline
\end{tabular}
\end{ruledtabular}
\end{table}

\begin{figure*}[!hbt]
    \begin{center}
\includegraphics[scale=0.43]{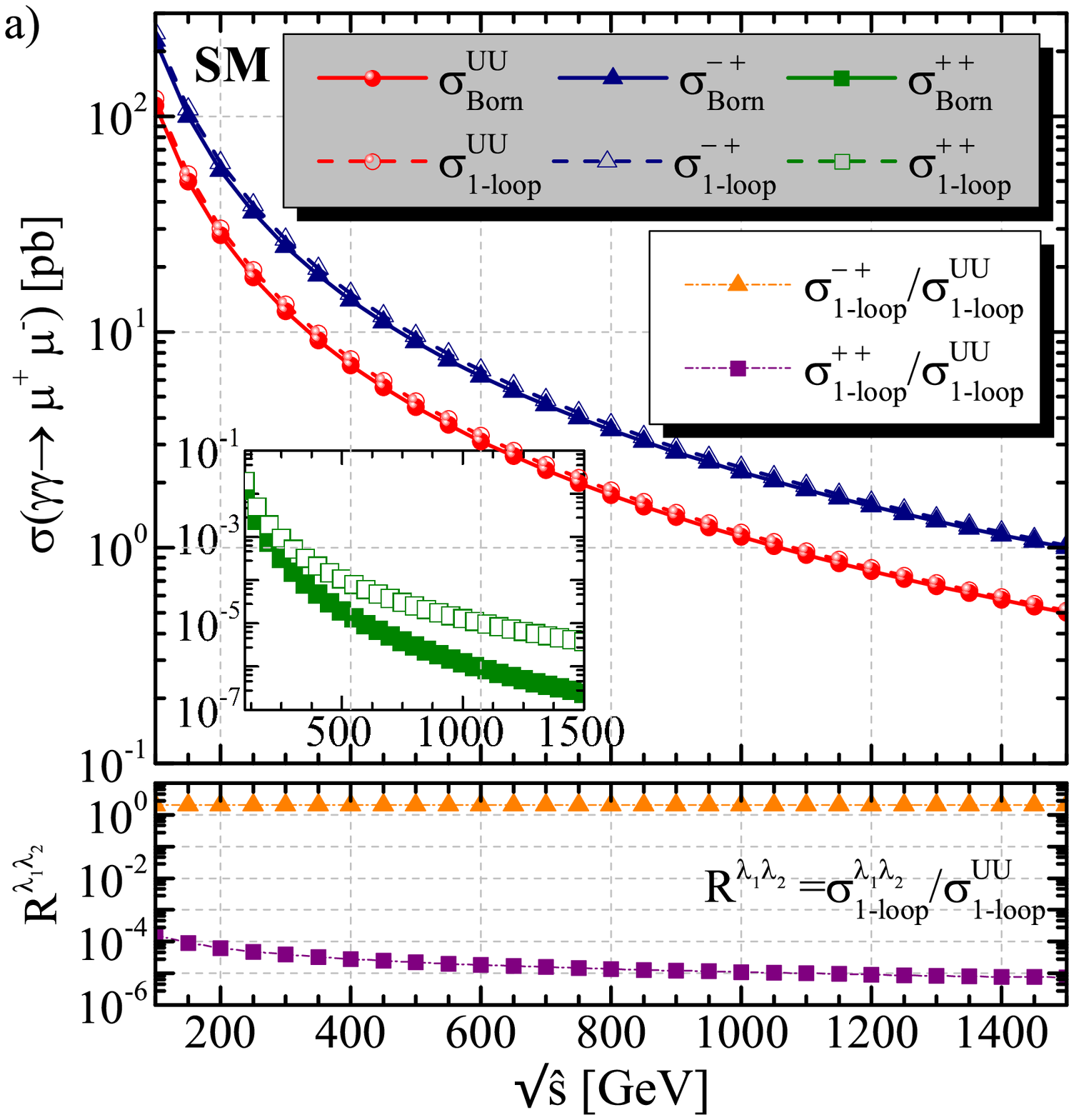}
\includegraphics[scale=0.43]{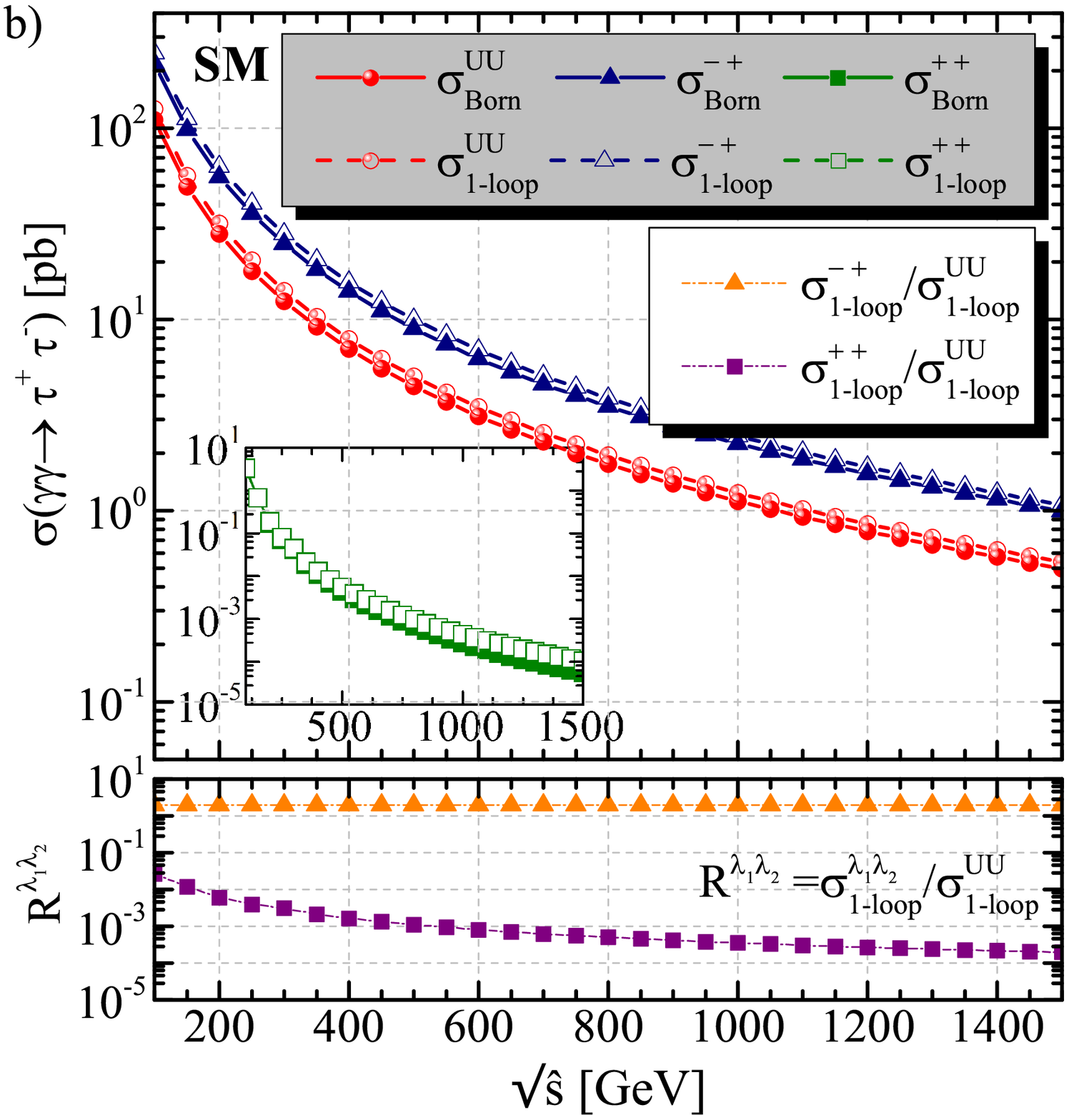}
     \end{center}
     \vspace{-4mm}
\caption{(color online). Polarized Born-level and one-loop cross sections of processes a) $\gamma\gamma \to \mu^- \mu^+$ and b) $\gamma\gamma \to \tau^- \tau^+ $ as a function of center of mass energy.}
\label{fig:pol}
\end{figure*}

In Fig.~\ref{fig:pol}(a)-(b), we show the initial beam polarization dependence of the Born-level and complete one-loop cross sections on center of mass energy for $\gamma \gamma \to \mu^- \mu^+$ and $ \gamma \gamma \to \tau^- \tau^+$. Also, we present the rate of $R^{\lambda_1\lambda_2}=\sigma^{\lambda_1\lambda_2}_{1-loop}/\sigma^{UU}_{1-loop}$ in order to see effect of polarization modes on the total cross section. The curves correspond to the integrated cross section with respectively the right-handed polarized photons $(++)$ with photon helicity $J_z = 0$, oppositely-polarized photons $(-+)$ with $J_z = 2$, and unpolarized photons $(\text{UU})$. Notice that the $(-+)$ and $(+-)$ polarized cross sections are the same, i.e. $\sigma^{-+}=\sigma^{+-}$.
\begin{figure*}[!hbt]
    \begin{center}
\includegraphics[scale=0.43]{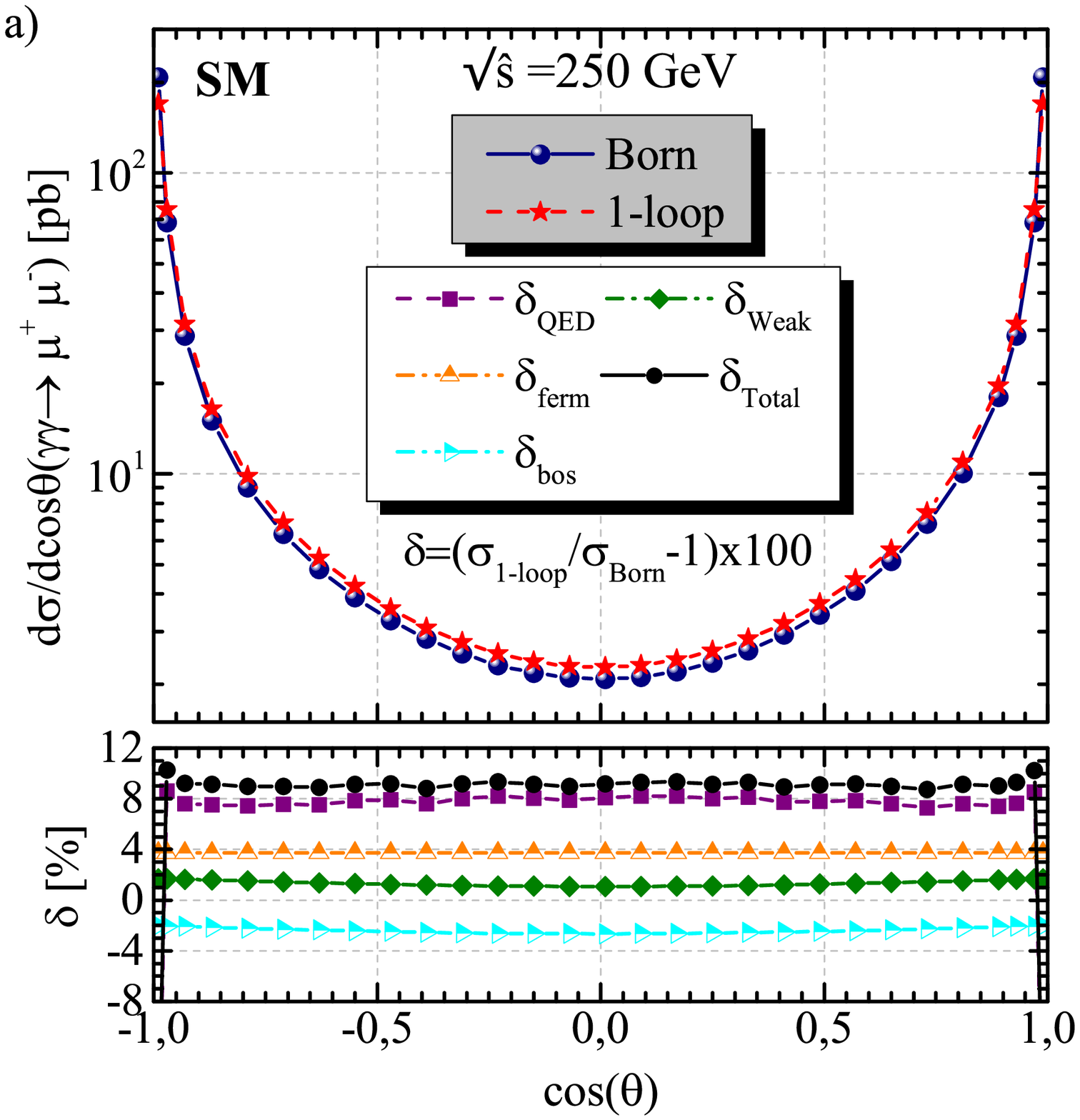}
\includegraphics[scale=0.43]{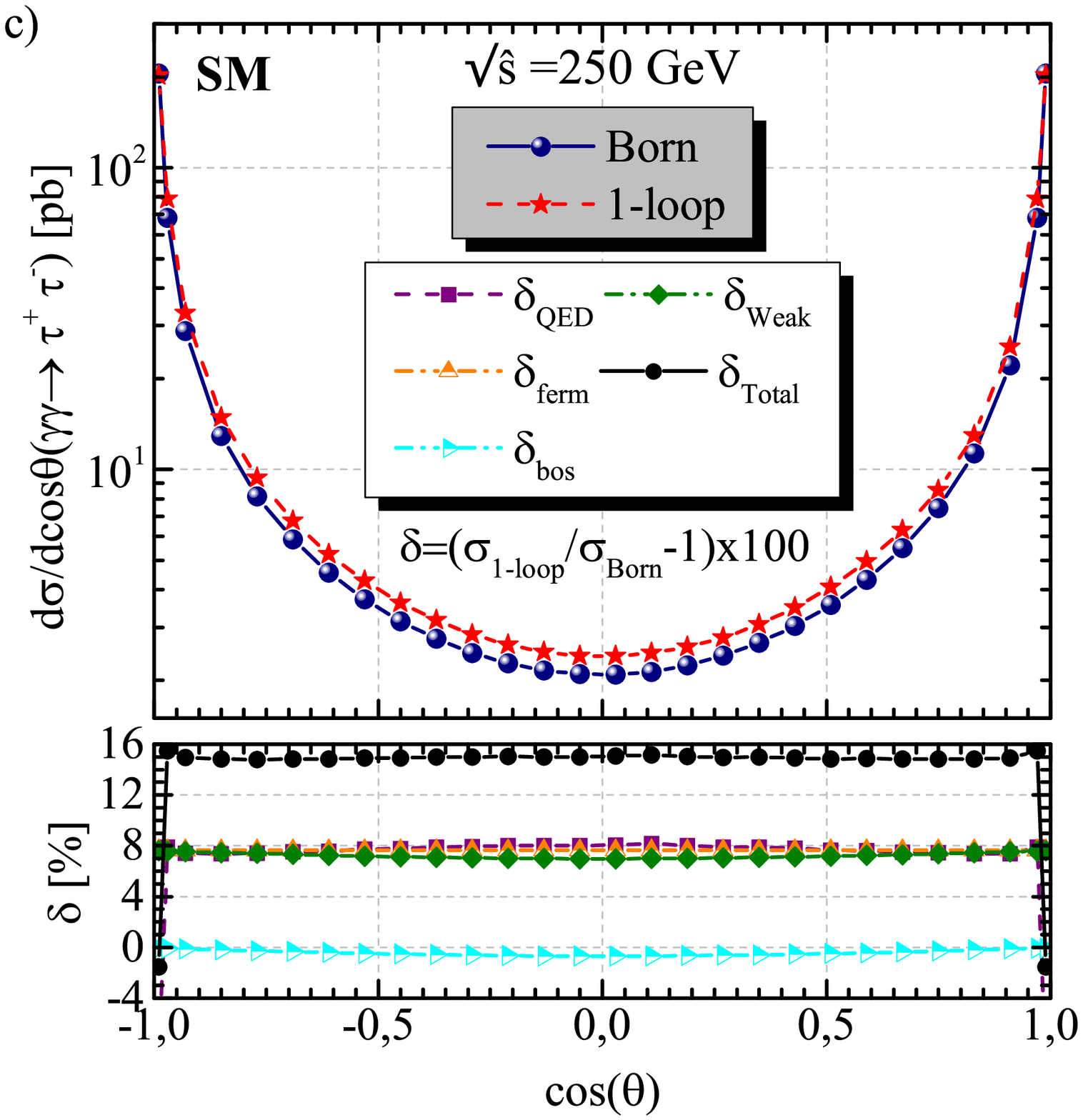}
     \end{center}
     \vspace{-4mm}
\caption{(color online). Angular distribution of the unpolarized Born-level and the one-loop differential cross sections for processes a) $\gamma\gamma \to \mu^- \mu^+$ and b) $\gamma\gamma \to \tau^- \tau^+ $ at $\sqrt{\hat{s}}=250\gev$.}
\label{fig:angular250}
\end{figure*}

From this figure, all curves for polarized and unpolarized cases start from their maximum values and fall off rapidly with the increment of center of mass energy. This is also the expected behavior. The cross sections with total photon helicity $J_z = 0$ ($\sigma^{++}$) turn on quickly, because they behave like $x$ at threshold,  whereas the $J_z = 2$ case ($\sigma^{-+}$) goes like $x^3$. Compared to the unpolarized case, the integrated Born and one-loop cross sections with oppositely polarized photons are increased by a factor of two. For the same-handed polarized photons, $(++)$ or $(--)$, the cross section is highly suppressed although at high energies. The rate $R^{\lambda_1\lambda_2}$ is around 2 and $10^{-4}$ for $\lambda_1\lambda_2=-+$ and $\lambda_1\lambda_2=++$, respectively. These behaviors show that having both photons polarized may be significant to provide a measurable production rate. For process $\gamma\gamma \to \mu^- \mu^+ $, at 100 GeV, $\sigma^{\text{UU}}_{\text{1-loop}}(\gamma\gamma \to \mu^- \mu^+ )$ have a maximum of 120.83 pb and the corresponding relative correction $\delta^{UU}_{\text{Total}}(\gamma\gamma \to \mu^- \mu^+ )$ is $+7.77\%$. The $\sigma^{+-}_{\text{1-loop}}(\gamma\gamma \to \mu^- \mu^+ )$ have a maximum of 242.56 pb, yielding total relative correction of about $+8.18\%$. On the other hand, for process $\gamma\gamma \to \tau^- \tau^+ $, $\sigma^{\text{UU}}_{\text{1-loop}}(\gamma\gamma \to \tau^- \tau^+ )$ have a maximum of 126.16 pb and the corresponding relative correction $\delta^{UU}_{\text{Total}}(\gamma\gamma \to \tau^- \tau^+ )$ is $+13.94\%$. The $\sigma^{+-}_{\text{1-loop}}(\gamma\gamma \to \tau^- \tau^+ )$ have a maximum of 249.48 pb and the corresponding $\delta^{+-}_{\text{Total}}(\gamma\gamma \to \tau^- \tau^+ )$ is $+14.25\%$.  Moreover, the absolute relative corrections decrease as $\sqrt{\hat{s}}$ rises. The full relative $\mathcal{O}(\alpha)$ correction is increased up to a few percent by oppositely polarized photons. As a result, the longitudinal polarization of initial beams increases the $\ell^- \ell^+$ production event rate in the photon-photon colliders.

\begin{figure*}[!htb]
    \begin{center}
\includegraphics[scale=0.43]{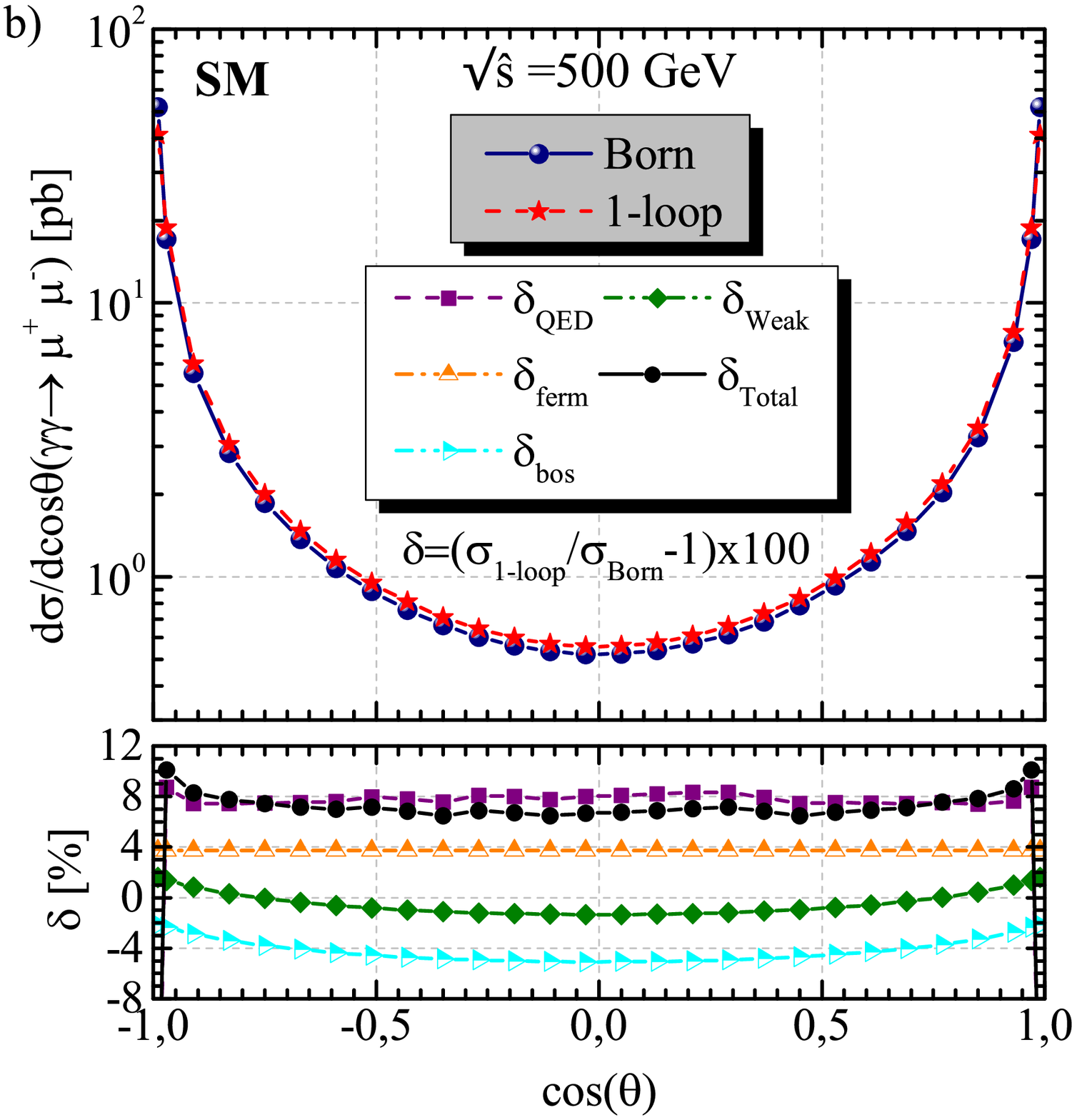}
\includegraphics[scale=0.43]{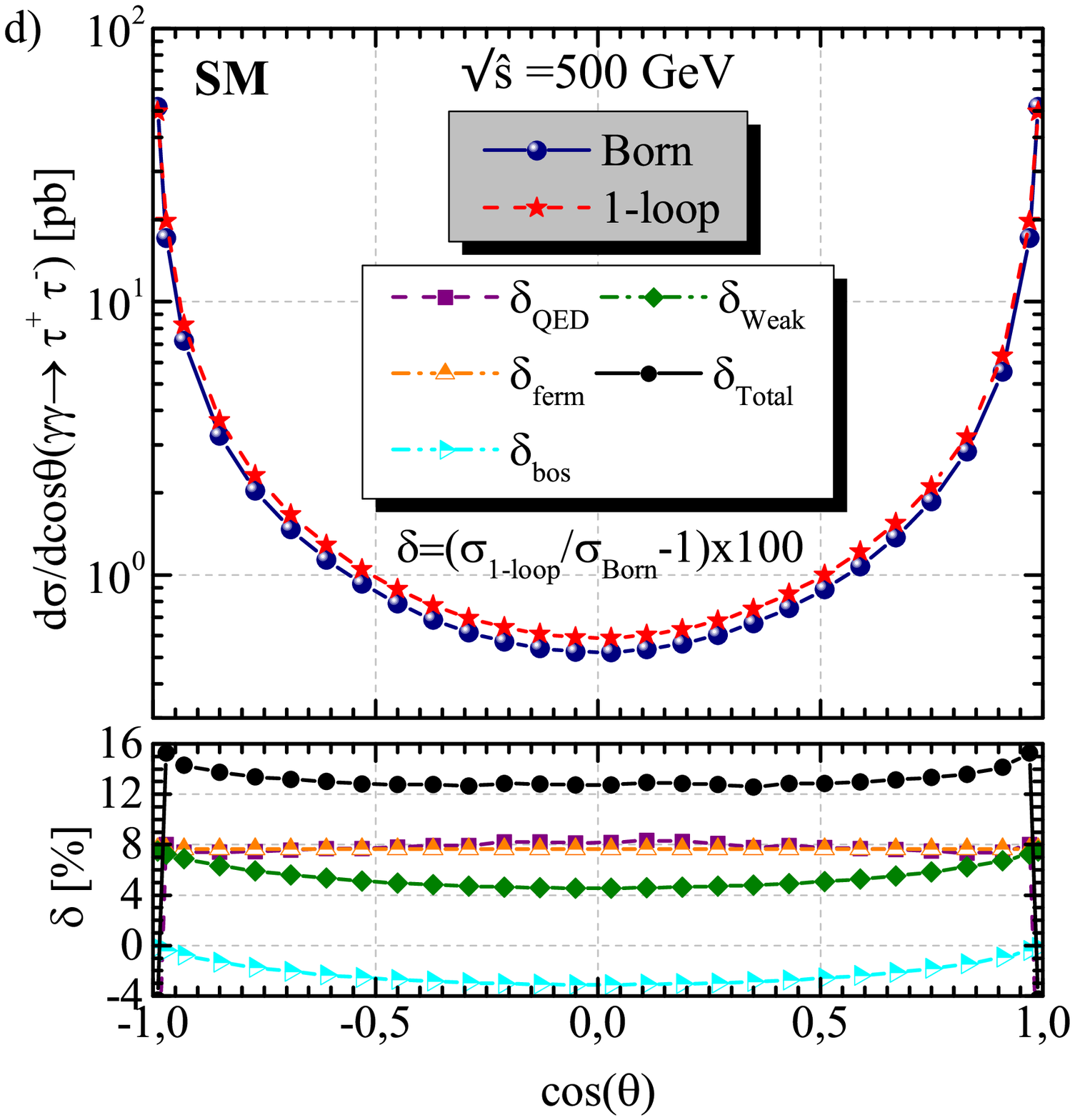}
     \end{center}
     \vspace{-4mm}
\caption{(color online). Same as in Fig.~\ref{fig:angular250}, but for $\sqrt{\hat{s}}=500\gev$.}
\label{fig:angular500}
\end{figure*}
In Figs.~\ref{fig:angular250} and~\ref{fig:angular500}, we present the Born-level and one-loop level of differential cross sections of $ \gamma \gamma \rightarrow \ell^{-}\ell^{+}$ as a function of the angle between the initial photon and the charged lepton at $\sqrt{\hat{s}}=250\gev$ and $500\gev$. We also provide the angular dependence of the relative effect of the EW corrections on the bottom panel of the same figures. The angular distribution of the unpolarized Born-level and the one-loop cross sections are (symmetrically) strongly peaked in the backward and forward directions. The relative corrections modify somewhat the Born-level angular distribution since their influence is larger in the central region. The relative corrections are relatively flat for $\sqrt{\hat{s}}=250\gev$, especially in the $-0.8 <\cos{\theta}<0.8$ region, yet the dependence on the angle becomes more distinctive for $\sqrt{\hat{s}}=500\gev$. The corrections reach their maximums as $\cos{\theta}$ approaches the extreme points; -1 or +1. Therefore, the charged leptons are dominantly produced in the forward and backward directions and it would be more probable to observe them in this collision region. By changing $\cos{\theta}$, the fermionic corrections stay almost stable, while the bosonic corrections supply a determinative contribution to weak corrections. Thus, these two corrections partially cancel as they combined forming the full weak corrections. However, QED contributions show a small dependence on $\cos{\theta}$. Furthermore, both QED and weak contributions are partially compensated with each other in the EW corrections. As $\cos{\theta}$ goes from 0 to +0.97 or -0.97, at $\sqrt{\hat{s}}=250\gev$, the full relative correction $\delta_{\text{Total}}$ varies from $9.1\%$ to $10.3\%$ for $\gamma\gamma \to \mu^- \mu^+ $, while $15.1\%$ to $15.5\%$ for $\gamma\gamma \to \tau^- \tau^+$. In the same $\cos{\theta}$ range for $\sqrt{\hat{s}}=500\gev$, the full relative correction $\delta_{\text{Total}}$ varies from $7.6\%$ to $10.1\%$ for $\gamma\gamma \to \mu^- \mu^+ $, while $12.7\%$ to $15.3\%$ for $\gamma\gamma \to \tau^- \tau^+$.

\begin{figure*}[!hbt]
    \begin{center}
\includegraphics[scale=0.43]{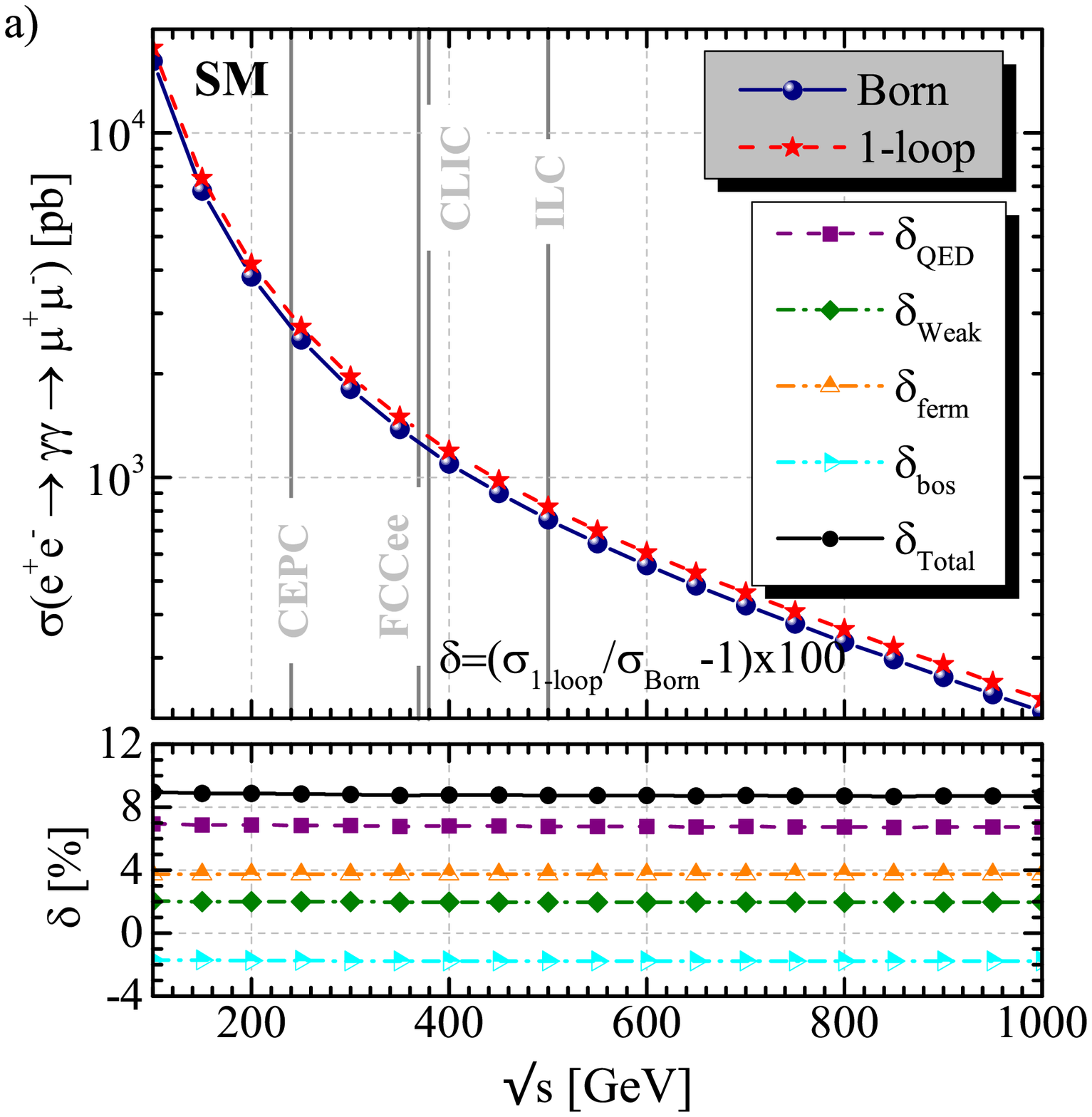}
\includegraphics[scale=0.43]{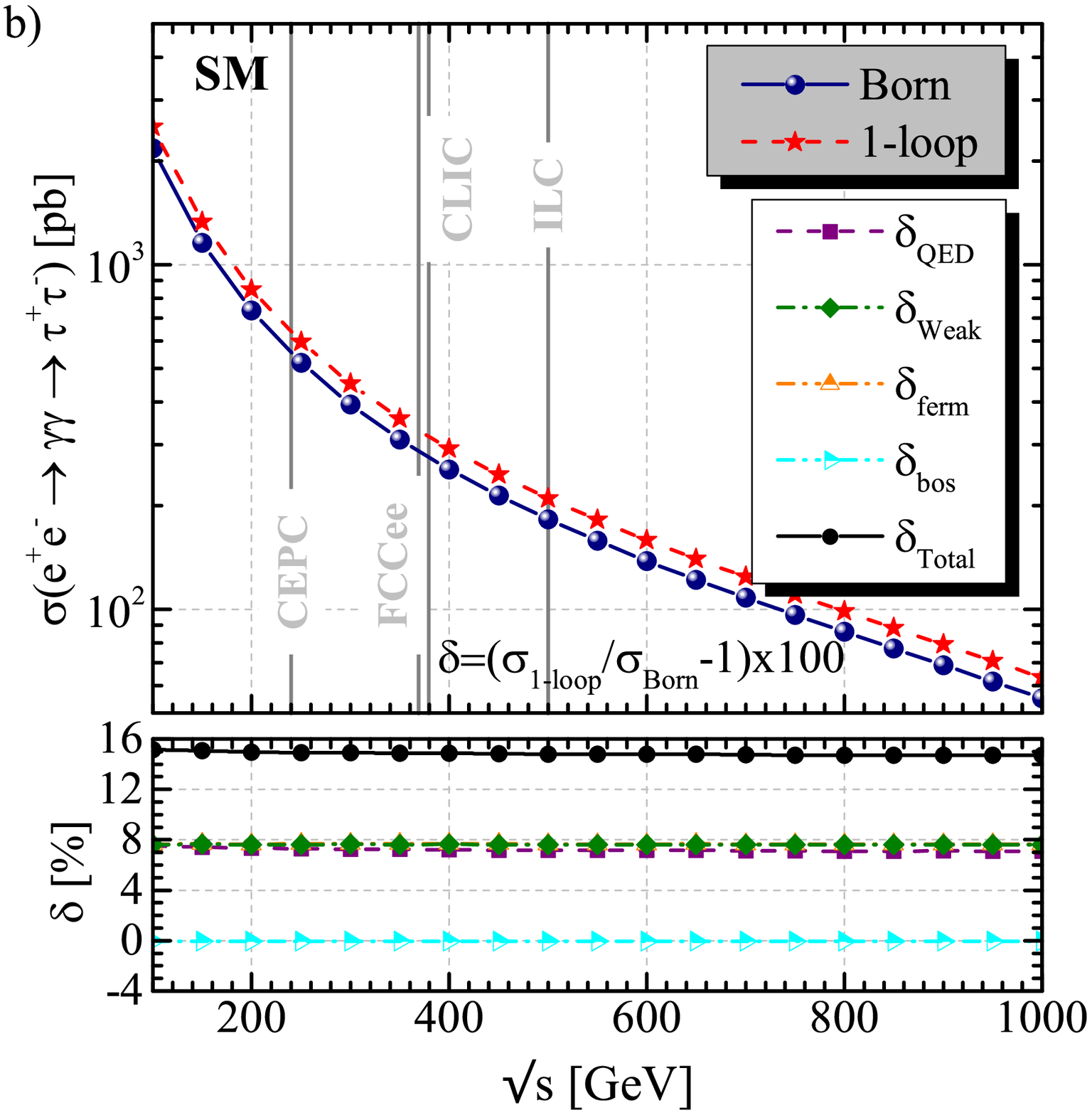}
     \end{center}
     \vspace{-4mm}
\caption{(color online). Born-level and full one-loop cross-sections convoluted with the photon luminosity of the parent processes a) $e^+e^-\rightarrow\gamma\gamma\rightarrow \mu^- \mu^+$ and b) $e^+e^-\rightarrow\gamma\gamma\rightarrow \tau^- \tau^+$ versus $\sqrt{s}$. The vertical solid-lines denote to the proposed energies of future colliders.
}
\label{fig:parentproc}
\end{figure*}
In Fig.~\ref{fig:parentproc}, we present the Born and full one-loop level cross sections for $e^- e^+ \rightarrow \gamma \gamma \rightarrow \ell^{-}\ell^{+}$ ($\ell=\mu,\tau$), calculated by convoluting with the photon luminosity, as a function of $e^- e^+$ center-of-mass energy. Their relative corrections that describe the effect of each correction type on the Born-level cross section as a function of $\sqrt{s}$ is also provided.  We can observe the expected behavior: a rapid decrease with the rise of the center-of-mass energy. The distributions have the same trend as in the subprocess. As the $\sqrt{s}$ goes from 100 GeV to 1 TeV, one-loop cross section decreases from 17.612 nb to 227.15 pb for $\gamma \gamma \to \mu^- \mu^+$ and 2.507 nb to 63.36 pb for $ \gamma \gamma \to \tau^- \tau^+$. The full EW corrections provide positive contribution to the total cross section in the parameter regions, and decrease with the increment of $\sqrt{s}$.

In Tables~\ref{table:parentprocmu} and~\ref{table:parentproctau}, we list numerical values for the unpolarized cross sections and the relative corrections for parent processes $ e^- e^+ \rightarrow \gamma \gamma \to \mu^- \mu^+$ and $e^- e^+ \to  \gamma \gamma \to \tau^- \tau^+$, respectively.
\begin{table}[!ht]
\caption{Born-level, one-loop cross sections (in pb) and relative corrections (in $\%$) for parent process $e^- e^+ \to \gamma \gamma \to \mu^- \mu^+$ at $\sqrt{s}=250\gev$, $500\gev$ and $1000\gev$.}\label{table:parentprocmu}
\centering
\begin{ruledtabular}
\begin{tabular}{lrrr}
%\hline
$\sqrt{s}$ & $250\gev$ & $500\gev$ & $1000\gev$\\
\hline
&\multicolumn{3}{c}{$\sigma(e^- e^+ \to \gamma\gamma \to \mu^- \mu^+)$ [pb]}\\
$\sigma_{\text{Born}}$   &2509.99  &755.58 &208.97 \\
$\sigma_{\text{1-loop}}$  &2731.52  &821.70 &227.15 \\
&\multicolumn{3}{c}{$\delta(e^- e^+ \to\gamma\gamma \to \mu^- \mu^+)$ [$\%$]}\\
$\delta_{\text{QED}}$   &+6.83  &+6.78 &+6.74 \\
$\delta_{\text{ferm}}$  &+3.75  &+3.74 &+3.74 \\
$\delta_{\text{bos}}$   &$-$1.75 &$-$1.77 &$-$1.79 \\
$\delta_{\text{Weak}}$  &+1.99  &+1.97 &$-$1.96 \\
$\delta_{\text{Total}}$ &+8.83  &+8.75 &+8.70 \\
\end{tabular}
\end{ruledtabular}
\end{table}
\begin{table}[!ht]
\caption{Same as Table~\ref{table:parentprocmu} but for $e^- e^+ \to \gamma \gamma \to \tau^- \tau^+$.}\label{table:parentproctau}
\centering
\begin{ruledtabular}
\begin{tabular}{lrrr}
%\hline
$\sqrt{s}$ & $250\gev$ & $500\gev$ & $1000\gev$\\
\hline
&\multicolumn{3}{c}{$\sigma(e^- e^+ \to \gamma\gamma \to \tau^- \tau^+)$ [pb]}\\
%\hline
$\sigma_{\text{Born}}$  &519.14  &182.62 &55.23\\
$\sigma_{\text{1-loop}}$ &596.58 &209.65 &63.36\\
%\hline
&\multicolumn{3}{c}{$\delta(e^- e^+ \to\gamma\gamma \to \tau^- \tau^+)$ [$\%$]}\\
$\delta_{\text{QED}}$   &+7.29  &+7.18 &+7.09 \\
$\delta_{\text{ferm}}$  &+7.68  &+7.67 &+7.67 \\
$\delta_{\text{bos}}$   &$-$0.00045 &$-$0.00049 &$-$0.00059 \\
$\delta_{\text{Weak}}$  &+7.63  &+7.62 &7.61 \\
$\delta_{\text{Total}}$ &+14.92  &+14.80 &+14.71 \\
\end{tabular}
\end{ruledtabular}
\end{table}

The fermionic and bosonic corrections are around $+3.7\%$ and $-1.8\%$ for $e^- e^+ \to \gamma \gamma \to \mu^- \mu^+$ and $+7.7\%$ and $-0.0005\%$ for $ e^- e^+ \to \gamma \gamma \to \tau^- \tau^+$, and depend only weakly on the center of mass energy. Therefore, the fermionic and the bosonic corrections partially cancel as they are combined to form the full weak corrections. The QED corrections are around $+6.95\%$ and $+7.55\%$ at the first point, which hardly changes up to 1 TeV for both processes. In the full EW corrections, both QED and weak corrections partially compensate each other, yielding relative corrections of about $+8.98\%$ and $+15.16\%$ at the first point, and $+8.70\%$ and $+14.71\%$ for $e^- e^+ \to\gamma \gamma \to \mu^- \mu^+$ and $ e^- e^+ \to\gamma \gamma \to \tau^- \tau^+$, respectively, at 1000 GeV.

\section{Summary and Conclusions}\label{sec:conc}
High precision calculations are required to further precision tests of the SM and search for clues on BSM. It is important to investigate such calculations in a simple and clean process such as the leptons production at $\gamma \gamma$ collisions that would provide an observable signal. A complete set of one-loop EW corrections must be included in production channels to ensure sufficient precision. In this study, we have studied the charged leptons pair production via $\gamma \gamma$ collisions, by considering a complete set of one-loop EW corrections. Accordingly, the UV divergences have been adjusted by dimensional regularization on the on-mass-shell renormalization scheme, while the IR divergences have been cancelled by the involvements of soft and hard QED radiation. The numerical evaluation was also made for two different schemes, the so-called the $\alpha(0)$-scheme and $G_\mu$-scheme. The $G_\mu$-scheme calculations are shifted by about $\%4$, comparing to the $\alpha(0)$-scheme. We have checked the stability of our result on the variation of the soft-hard cutoff parameter. We also compared the Born-level and the hard photon bremsstrahlung cross-sections with the results of \textsc{CalcHEP} and \textsc{Whizard} and found very good (within four to five digits) agreement with the packages used in this work.

The results indicate that the one-loop EW radiative corrections mostly improve the Born-level cross section and the total relative correction is typically about ten percent for both $\gamma\gamma\rightarrow \mu^{-}\mu^{+}$ and $\gamma\gamma\rightarrow \tau^{-}\tau^{+}$  processes. We have also investigated the interplay between weak fermionic, bosonic, and QED corrections. The fermionic corrections depend weakly on the center of mass energy, whereas the weak bosonic contribution is negative and decreases rapidly with increasing center of mass energies. Therefore, the fermionic and bosonic contributions partially cancel when combined to form the full weak corrections. Additionally, the QED corrections do not depend on the center of mass energy for $s\gg m^2_\ell$, because all mass singularities cancel, and $s$ is the only scale that survives. The QED plus weak corrections are partially compensated each other, yielding relative corrections of about ten percent.

Moreover, we have presented numerical results for angular distribution and the initial beam polarisation dependence of the Born-level and one-loop cross sections. As a result, the angular distribution is (symmetrically) strongly peaked in the backward and forward directions. The relative corrections modify somewhat the Born-level angular distribution since their effect is larger in the central region. On the other hand, it is obvious from polarization distribution that the polarization effects are significant and enhance the cross section at the definite initial degrees of polarization as compared to the unpolarized one.

In summary, the detailed phenomenological results for the one-loop EW radiative corrections to muon-pair and tau-pair productions through $\gamma\gamma$ collisions have been presented in the framework of the SM. In the light of this, the effects on the cross-section of relative contributions created by dividing virtual contributions into the gauge-invariant subsets have been separately investigated. One-loop EW radiative corrections are clearly shown to change the lowest-order results significantly, and must thus be taken into complete account for a realistic description of experiments at future collider energies. Our results will be helpful for matching a percent level accuracy.

\begin{acknowledgments}
A part of the computations reported in this study was performed at the National Academic Network and Information Center (ULAKBIM) of TUBITAK, High Performance and Grid Computing Center (TRUBA Resources).
\end{acknowledgments}

\end{document}